\documentclass[prl,reprint,aps,superscriptaddress,longbibliography,floatfix]{revtex4-2}

\newcommand{\be}{\begin{equation}}
\newcommand{\ee}{\end{equation}}

\newcommand{\bea}{\begin{eqnarray}}
\newcommand{\eea}{\end{eqnarray}}

\newcommand{\abs}[1]{\left| #1 \right|} 
 
\newcommand{\ket}[1]{\left| #1 \right>} 
\newcommand{\bra}[1]{\left< #1 \right|} 

\let\baraccent=\= 

\usepackage{xcolor}
\usepackage{graphicx}
\usepackage{dcolumn}
\usepackage{bm}
\usepackage{amsmath}
\usepackage{placeins}
\usepackage{verbatim}
\usepackage{dsfont}

\begin{document}

\preprint{APS/123-QED}

\title{Majorana edge modes in isolated wires}

\author{Jaden Thomas-Markarian}
\email{jthomasm@mit.edu}

\affiliation{Department of Physics, Massachusetts Institute of Technology, Cambridge, MA 02139, USA}
\affiliation{Department of Physics, University of Chicago, Chicago, Illinois 60637, USA}

\author{Kartiek Agarwal}

\email{kagarwal@anl.gov}

\affiliation{Materials Science Division, Argonne National Laboratory, Argonne, IL 60439, USA}

\author{Ivar Martin}

\email{ivar@anl.gov}

\affiliation{Department of Physics, University of Chicago, Chicago, Illinois 60637, USA}
\affiliation{Materials Science Division, Argonne National Laboratory, Argonne, IL 60439, USA}

\date{\today}

\begin{abstract}
Topological superconductors are believed to host exotic quasiparticle excitations known as Majorana zero-modes (MZMs), with much of the evidence based on BCS mean-field theory. The direct application of mean-field arguments is tenuous in finite, isolated systems relevant in some experiments. Here, we develop a new correlation-based method for identifying MZMs in interacting, number-conserving systems. Using DMRG, we study fermion number-conserving  models with long-range interactions, which under periodic boundary conditions exhibit robust topological and non-topological superconductivity, tuned by the strength of interaction \cite{ortiz2014many}. We find evidence that, on the topological side, Majorana edge modes appear in open chains, manifesting as the vanishing of the energy splitting between odd- and even-parity ground states with increasing system size. Additionally, off-diagonal two-point correlation functions show nonlocal, parity-dependent edge effects. These correlations reveal the spatial structure of Majorana modes in the many-body wavefunction. We show that the correlation diagnostic applies broadly, including to short-range interacting models, where topological superconductivity is more fragile due to the absence of a bulk excitation gap.
\end{abstract}

\maketitle


In the last few decades, the physics of Majorana fermions has become an important subfield of condensed matter physics. Majorana fermions were first constructed as real solutions to the relativistic Dirac equation~\cite{majorana1981symmetric}. Formally, any complex fermion can be written as a superposition of two Majorana fermions. However, only in special cases, when a system exhibits topological order, can spatially unpaired Majorana fermions exist. These operators are self-conjugate and commute with the Hamiltonian. Hence, they represent zero-energy quasiparticles and are referred to as Majorana zero-modes (MZMs). MZMs have garnered significant theoretical and experimental interest in recent years due to their potential applications in fault-tolerant quantum computing and topologically protected quantum information storage~\cite{kitaev2003fault, nayak2008non}. MZMs possess unique properties stemming from their non-Abelian exchange statistics and topological protection that make them robust against local perturbations and decoherence. Their potential realization in condensed matter systems such as topological superconductors has sparked intense experimental~\cite{finck2013anomalous, churchill2013superconductor,nadj2014observation,wang2018evidence,aghaee2023inas} and theoretical efforts~\cite{pikulin2021protocol,vijay2016teleportation,martin2020double,min2022dynamical,truong2023optimizing,truong2025shuttling,Pientka2017,Fu2008,Lutchyn2010,Alicea2012,NadjPerge2013,Yang2016,Martin2012,https://doi.org/10.48550/arxiv.2602.09156} aimed at their detection and manipulation.

The canonical example of a system supporting MZMs is the Kitaev chain~\cite{kitaev2001unpaired}. Kitaev's model is a mean-field description of a $1$D tight-binding chain 
with p-wave superconducting pairing. 
It supports a topological phase 
characterized by the presence of unpaired MZMs localized near the chain's edges. 
Kitaev's model forms the basis for understanding more realistic models of topological states in superconducting wires, with spinful electrons, noise and disorder~\cite{alicea2012new}. However, being a mean-field model, it lacks particle number conservation, which should be an important consideration for describing finite,  isolated segments of superconductors or superfluids. 
Significant effort has thus been dedicated to understanding systems that support MZMs and conserve particle number~\cite{fidkowski2011majorana, Sau2011, Cheng2011, vodola2014kitaev, iemini2015localized, Lang2015, wang2017number, lin2018towards,  knapp2020number, sajith2024signatures, martin2024majorana}. 

Most of the considered cases included only short-range interactions, which can only sustain finite-range superconducting correlations in 1D. In contrast, long-range interactions can stabilize phase coherence over long distances even in $1$D. This has been demonstrated by Ortiz et al~\cite{ortiz2014many}, using integrable long-range interacting Richardson-Gaudin-Kitaev (RGK) models on a ring. They convincingly demonstrated the existence of topological and non-topological superconducting phases by probing ground state fermion parity switches between periodic and antiperiodic boundary conditions. However, the integrability breaks down for open boundary conditions and hence cannot be used to test for the existence of edge modes.

In this letter, we develop a new correlation-based approach for characterizing Majorana edge modes in interacting, number-conserving systems and implement it numerically using DMRG. As a controlled setting in which the thermodynamic-limit properties are well understood~\cite{ortiz2014many}, we first apply this framework to the RGK model on open segments.
First, we examine the odd-even fermion parity energy gap, $\Delta E$, in the open chain as a function of system size $L$. In the trivial phase, this difference saturates to a finite value. In the topological phase, it decays to zero as a power law $\sim 1/L$. The latter behavior can be attributed to the presence of gapless edge modes. Second, we find that single-fermion two-point correlation functions clearly distinguish the topological phase from the trivial phase. We present a correlation-based method to extract the real-space profiles of excitations that correspond to the parity switches. We find that in the topological case, these excitations have mixed particle-hole character, in close analogy to their mean-field counterparts. That stands in contrast to the trivial case, where parity-switching excitations have predominantly particle character, as one would expect in the case of single-particle excitations in the BEC limit of a superconductor. 
Finally, we calculate observables that can be associated with the parities of Majorana edge modes and find that they are quantized, in direct correspondence with mean-field MZMs



\emph{Richardson-Gaudin-Kitaev Chain.} 
The Hamiltonian defining the RGK chain~\cite{ortiz2014many} is
\begin{align}
    \mathcal H_{\textnormal{RGK}} = \sum_{k} \varepsilon_k c_k^{\dagger}c_k - \frac{16g}{L}\sum_{k,k'}\eta_k \eta_{k'}c_k^{\dagger}c_{-k}^{\dagger}c_{-k'}c_{k'}, \label{eq:H_rkg}
\end{align}
where $\varepsilon_k = -2t_1\cos{k} - 2t_2\cos{2k}$ is the single-particle dispersion, $g>0$ is the interaction strength, and $c_k$ are annihilation operators for spinless particles with momentum $k$. 
The special form of the interaction term 
$\eta_k = \sin{(k/2)}\sqrt{t_1 + 4t_2\cos^2{(k/2)}}$
makes the model integrable, allowing for exact numerical solution for large system sizes 
both for periodic and antiperiodic boundary conditions. 
General boundary conditions are encoded by the phase $\phi$, $c_{i+L} = e^{i\phi}c_i$, with $\phi = 0$ corresponding to periodic and $\phi = \pi$ anti-periodic.
In real space, the Hamiltonian takes the form
\begin{align}
    \mathcal H_{\textnormal{RGK}} = -\sum_{i=1}^{L} \sum_{l=1}^{2} (t_l c_i^{\dagger}c_{i+l} + h.c.) - \frac{4g}{L} I_{\phi}^{\dagger} I_{\phi}, \label{eq:Hammy}
\end{align}
where 
\begin{align}
    I_\phi = 2i\sum_k \eta_k c_k c_{-k} = \sum_{i>j}^L \eta(i-j)c_ic_j.
\end{align}
Here we focus on the case $t_1 \neq 0$, $t_2 = 0$ (nearest neighbor hopping), and set $t_1 = 1$. The position space pairing potential $\eta(m)$ can be determined in closed form via Fourier transform, which in the limit $L \rightarrow \infty$ yields
\begin{align}
    \eta(r) = \frac{8r(-1)^r}{\pi(1-4r^2)}.
    \label{eq:iner}
\end{align}
Such long-range interactions can stabilize topological phases that tend to be fragile in systems with only short-range interactions in 1D~\cite{gong2016topological, choy2011majorana}.

\emph{Signatures of topological superconductivity and Majorana zero-modes.} Within mean-field descriptions of topological superconductors, one can easily compute bulk topological invariants and directly verify the existence of MZMs. 
The situation is more subtle for interacting systems.
In the closed chain, the topological phase transition can be identified from ground state fermion parity switches. The trivial phase is associated with even ground state parity: all electrons prefer to form Cooper pairs and adding an extra particle costs energy. In the topological phase, on the contrary, the ground state parity is {\em odd} for periodic boundary conditions. Notably, changing the boundary conditions from periodic to anti-periodic has no effect on the ground state parity in the trivial phase (even), while switching it to even in the topological case. In the mean-field description of Read and Green \cite{read2000paired}, the phases track the occupancy of the zero-momentum fermion  mode (empty in the trivial phase and occupied in the topological phase). This mode only exists for periodic boundary conditions and thus the spectral distinction between the trivial and topological phases disappears for anti-periodic boundary conditions.

In number-conserving models, it is also possible to track fermion parity switches under twisted boundary conditions~\cite{ortiz2014many}. However, care needs to be taken to eliminate the trivial density-dependent contribution to the energy, which can mask the effect. This can be done by considering the {``symmetrized"} odd-even ground state energy difference
$\Delta E = E_0^{\textnormal{odd}}(N,L) - E_0^{\textnormal{even}}(N,L)$. For even $N$,  we define $E_0^{\textnormal{odd}}(N,L) := \frac{1}{2}[E_0(N+1,L) + E_0(N-1,L)]$ and $E_0^{\textnormal{even}}(N,L) := E_0(N,L)$. For odd $N$, we swap the definitions of $E_0^{\textnormal{odd}}(N,L)$ and $E_0^{\textnormal{even}}(N,L)$. Averaging over $N\pm 1$ offsets the trivial (chemical potential) shift, allowing a comparison of energies with the same average density regardless of the parity sector. Ortiz et al~\cite{ortiz2014many} studied $\Delta E$ under periodic and anti-periodic boundary conditions to identify the topological phase transition in these models; we verify these findings in SM.I as a consistency check. 

Even when a bulk topological invariant can be defined, it may not guarantee the existence of gapless edge modes in the open chain~\cite{fidkowski2011topological, manmana2012topological}.
Instead, to directly probe the presence of MZMs in the open system, we analyze the symmetrized odd-even energy gap and the ground state single-particle two-point correlation matrix. As we show, these reveal qualitative differences between the trivial and topological phases.

\emph{Open RGK chain: Spectrum.}
A telltale sign of gapless edge modes is the behavior of the odd-even excitation gap, $\Delta E$, as a function of system size $L$. In the topological phase, the energy cost to change parity sectors is expected to vanish in the thermodynamic limit.  In contrast, in the trivial phase, no  low-energy edge quasiparticles exist, and 
$\Delta E$ should remain finite in the thermodynamic limit. \begin{figure}
\includegraphics[scale=0.40]{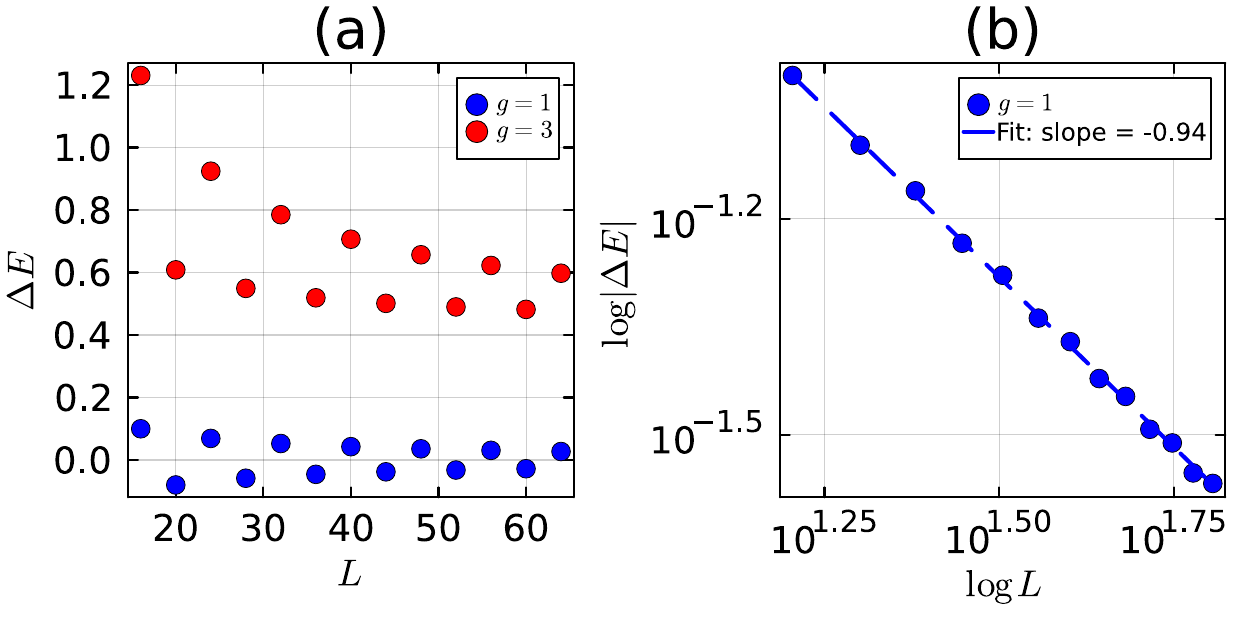}
\caption{(a) Odd-even excitation gap vs system size for open boundary conditions and $g=1$ (topological), $g=3$ (trivial). In the topological phase, the gap decays to zero, signaling the presence of gapless edge modes. (b) Log-log plot of odd-even excitation gap vs system size for $g=1$; the gap decreases as $\sim 1/L$.
}
\label{fig:E_gap2}
\end{figure}

In Fig.~\ref{fig:E_gap2}, we present numerical results for the topological and trivial phases. We choose the reference particle number as $N = L/4$ for each system size; it can be even or odd. In the RGK model, there is no phase transition at $1/2$ filling, whereas $g_c\approx2$ for $1/4$ filling. We find that in the topological phase, for $g = 1$, $|\Delta E|$ decreases as $\sim1/L$, consistent with the presence of gapless edge modes.  We note, however, that this $\Delta E$ does not have an interpretation as the quasiparticle energy cost; instead, the decay of $\Delta E$ should  be attributed to a smooth `charging energy' $E_0(N)\sim N^2/L$. Indeed, the energy cost to go from $N$ to symmetrized $N\pm 1$ states is always positive, regardless of whether $N$ is even or odd, consistent with the ground state energy being a \textit{nearly} smooth function of $N$ with vanishing discrete curvature $E_0(N+1) - 2E_0(N) + E_0(N-1)$ for large $L = 4N$. There is, however, a much smaller underlying $(-1)^N$ energy oscillation, which can be associated with the MZM energy splitting.  It can be extracted by taking higher discrete derivatives of the energy (SM.VI), and also found to scale as  $ 1/L$. 

On the other hand, in the trivial phase, the splitting $\abs{\Delta E}$ does not decay to zero, with the odd parity state always being higher in energy. This is consistent with the finite energy cost associated with an unpaired quasiparticle, as expected in a trivial superconductor~\cite{bardeen1957theory}. 

\emph{Open RGK chain: Correlation matrix.}
We next examine the single-particle density matrix $M^{ij} = \bra{N} c_i^\dagger c_j \ket{N}$, which, as we will show,
can directly reveal the structure of MZMs. In the ground state of a non-interacting system, the eigenmodes (and eigenvalues) of $M$ coincide with the single-particle eigenstates (and their occupation). In general, the eigenvalues are bounded between 0 and 1. Tracking changes in $M^{ij}$ between ground states at different $N$ can thus shed light on the structure of the lowest energy quasiparticle excitation connecting these states. In atomic simulators, $M^{ij}$ can be experimentally accessed via quenches~\cite{denzler2024learning}. 

\begin{figure} 
\includegraphics[scale=0.4]{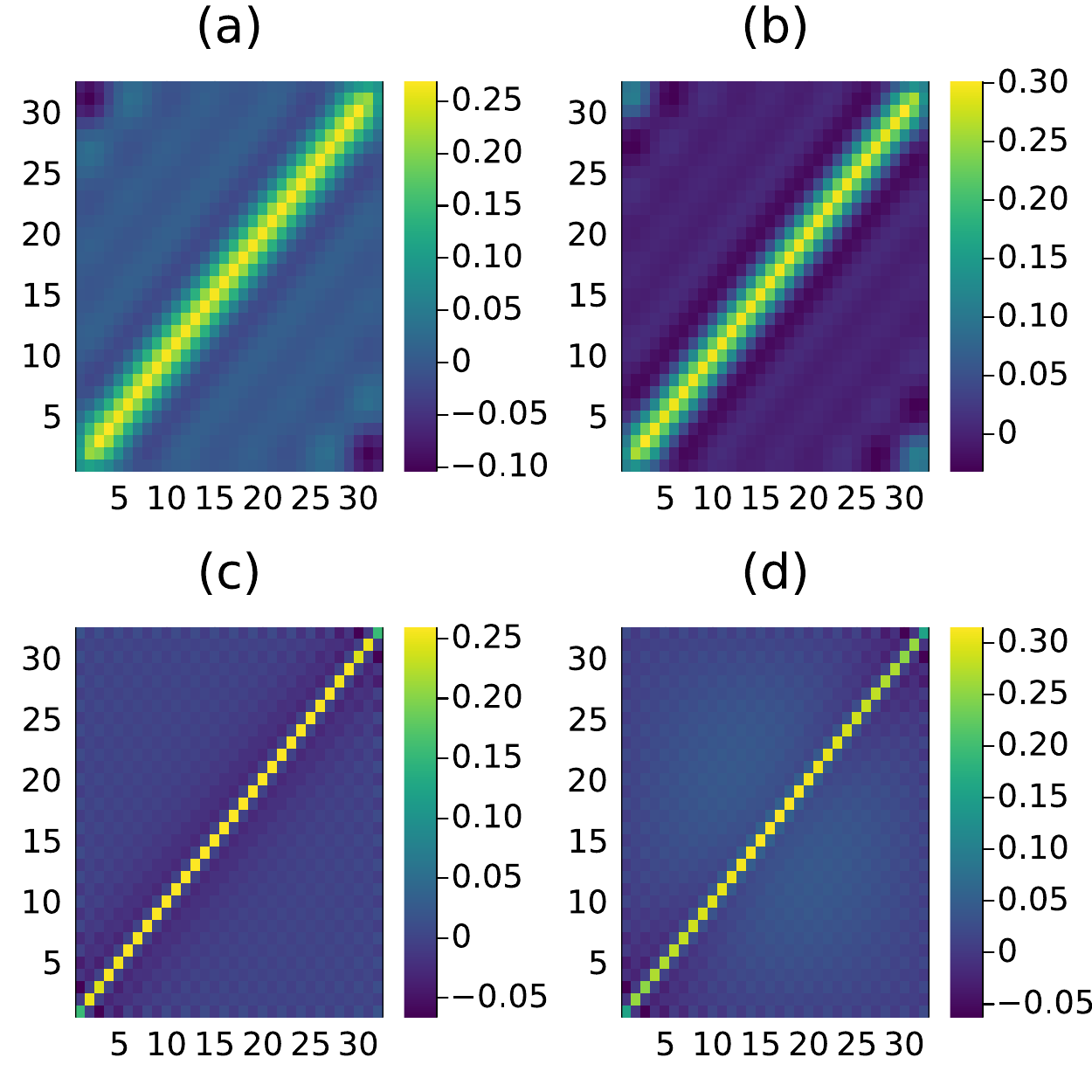} 
\caption{Heat maps of the two point correlation matrix $\langle c_i^\dagger c_j\rangle$ for (a) $g=1$, $L=32$ and $N=8$; (b) $g=1$, $L=32$ and $N=9$. The off-diagonal (inter-edge) correlators are large and differ in sign (comparing $N=8$ to $N=9$). For $g=3$, there are no obvious differences between (c) $N=8$ and (d) $N=9$.}
\label{fig:corr_mats}
\end{figure}

In Figure \ref{fig:corr_mats}, $M^{ij}$ is plotted for a system of size $L = 32$ in four cases: $ N=8$ and $N=9$, for $g=1$ and $g=3$. For the trivial case ($g = 3$), there is no obvious difference between $N = 8$ and $N=9$. In contrast,
for the topological case, the correlators involving creation/annihilation operations at the opposite edges 
are large and differ in sign for $N = 8$ and $N = 9$. This behavior remains robust as the system size increases (comparing $N=L/4, L/4+1$), as we verify in SM.VII. 
The finiteness of these inter-edge correlators at large $N$ suggests the presence of a topologically protected edge mode. 

\begin{figure}
\includegraphics[scale=0.4]{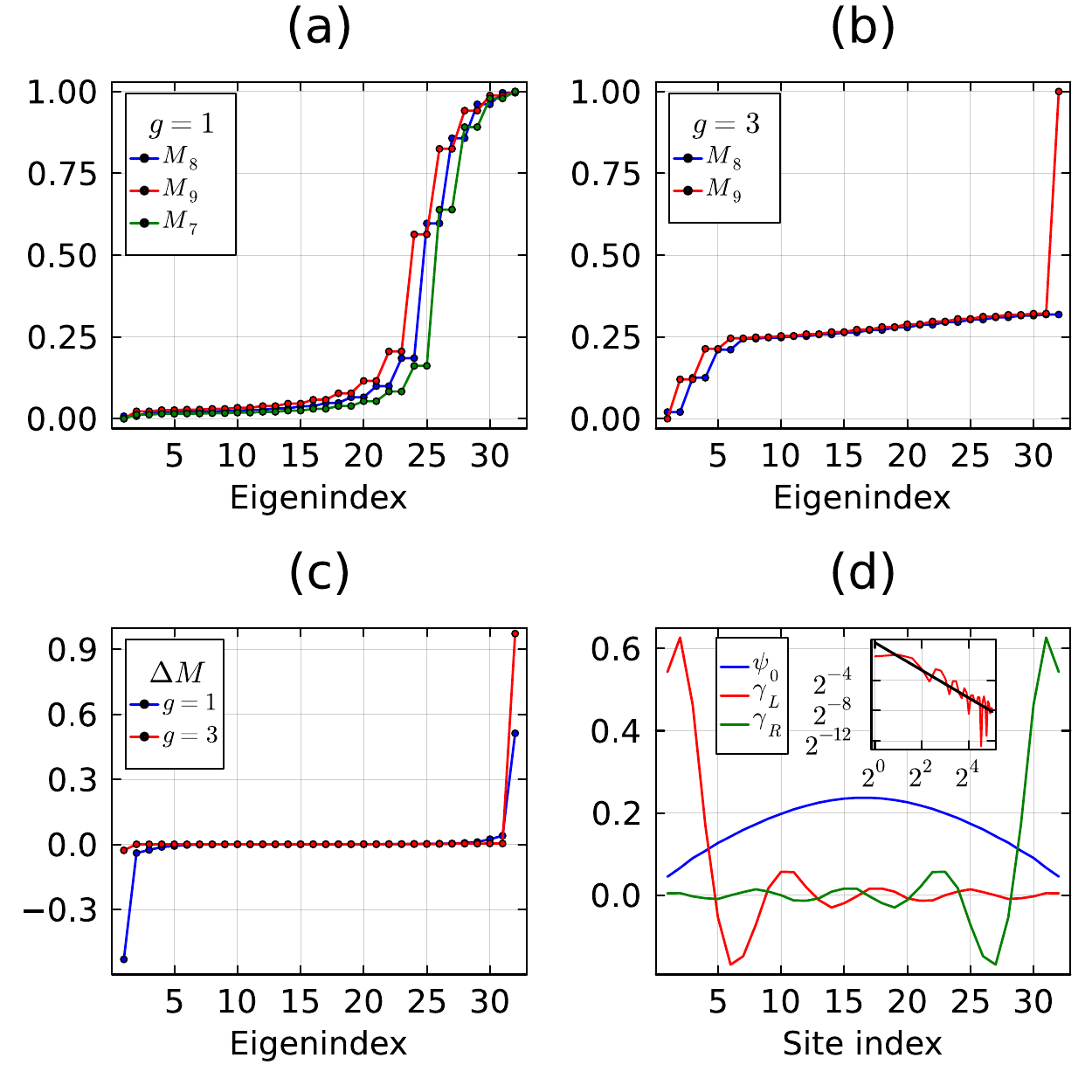}
\caption{ 
Spectra of correlation matrices (a) $M_7, M_8$ and $M_9$ in the topological phase; (b) $M_8, M_9$ in the trivial phase.  (c) The spectra of the odd-even difference matrix $\Delta M$ for the topological and trivial cases, defined in text.  (d) The eigenstate of $\Delta M$ corresponding to the eigenvalue 1 in the trivial case (blue), along with the sum (green) and difference (red) of the eigenstates with eigenvalues $\approx \pm 0.5$ in the topological case. The former corresponds to the $k \approx \pi/L$ mode,
while the latter can be  interpreted as the wavefunctions of the left and right MZMs $\gamma_L, \gamma_R$, respectively (see text). The inset shows $\gamma_L$  decays as a power-law $\sim 1/r^2$ away from the edge.}
\label{fig:spectra}
\end{figure}

The eigen-spectra of these matrices in the trivial and topological phases are shown in Fig.~\ref{fig:spectra} (a) and (b). We also define a new matrix $\Delta M \equiv M_o - M_e$, constructed out  of $M_e \equiv M_8$, and $M_o =[ M_7 + M_9]/2$  in the topological phase and simply $M_o = M_9$ in the trivial phase. The different choices of $M_o$ in the two phases lead to the smallest rank of $\Delta M$. We observe that i) the spectra for even and
odd parity states are doubly degenerate except the odd parity states have additional, isolated and nearly exact $0$ and $1$ eigenvalues, independent of the phase; ii) the topological phase is characterized by a spectrum spanning the
full range from 0 to 1, unlike the trivial phase where all the eigenvalues (except the eigenvalue 1 for the odd parity state) lie between $0$ and $0.35$;
iii)  $\Delta M$ is a low-rank matrix. In the trivial phase, all eigenvalues of  $\Delta M$ are close to $0$, except one, which is $\approx1$. In the topological phase, there are two large eigenvalues $\approx \pm 1/2$, with  eigenvectors whose weight is peaked near the edges. 

The pronounced eigenvalue doubling 
is evidence of Cooper pairing. As we show in SM.III, a number-projected BCS wavefunction~\cite{leggett2006quantum}, $\psi \sim P_N e^{A_{ij} c^\dagger_i c^\dagger_j} \ket{0} $, has this feature for even $N$. The skew-symmetric pairing matrix $A_{ij}$ can be diagonalized into blocks $\lambda_l (-i\sigma^y)$ associated with the $l^{\text{th}}$ pair of orthonormal modes.  The correlation matrix is diagonalized by the same orthonormal modes, but with eigenvalues $\lambda_l^2/(\lambda_l^2 + 1)\equiv n_l$, corresponding to the probability of a pair being present in the condensate. The difference in the span of eigenvalues between the two phases agrees with a mean-field 
treatment for a periodic system. The full span in the topological phase can be related with the Zak phase acquiring a nontrivial winding (see SM.III for details). 

The simplest prescription to create a charged excitation is to unbind a pair of modes with eigenvalue $\lambda_*$, and set the occupation of one mode in the pair to $1$, and the other to $0$. Up to $1/N$ corrections, this would leave the  spectrum of $M$ unchanged, except for the pair of eigenvalues $n_*$ converting into 1 and 0. Consequently,  with the same accuracy, $\Delta M$ would be a rank-2 matrix with eigenvalues $1-n_*$ and $-n_*$. 
In the trivial phase, the data is well explained with the choice $\lambda_* \approx 0$, both for open and periodic boundary conditions, indicating that the excitation is essentially particle-like. In the topological phase, the $\pm 1/2$ eigenvalues of $\Delta M$ imply $\lambda_* \approx 1$. This indeed works well for periodic boundary conditions, where a pair with  eigenvalues $n_*\approx 1/2$ are converted into 1 and 0 upon addition of a particle (see SM.III). 
However, for open boundary conditions in the topological phase, there is no evidence of such an unbinding of a selected pair with $\lambda_* \approx 1$ (see spectra of $M_7, M_8, M_9$ in Fig.~\ref{fig:spectra} (a)), motivating a more general ansatz.

\emph{Particle-hole symmetric ansatz.} Inspired by mean-field considerations, we conjecture the following relations between ground states in the topological phase with open boundary conditions:
\begin{align}
     \ket{N+1}_o & = \frac{1}{\sqrt{2}} \sum_j\alpha_j c^\dag_j\ket{N}_e + \frac{1}{\sqrt{2}}  \sum_j \beta_j c_j\ket{N+2}_e,  \nonumber \\
    \ket{N}_e & = \frac{1}{\sqrt{2}} \sum_j \beta_j c^\dag_j \ket{N-1}_o + \frac{1}{\sqrt{2}} \sum_j \alpha_j c_i\ket{N+1}_o,
    \label{eq:ansE}
\end{align}
    with normalization $\sum_j \abs{\alpha_j}^2 = \sum_j \abs{\beta_j}^2 = 1$. In the mean-field case, as well as in the number-projected Kitaev wave function~\cite{sajith2024signatures}, the excitation connecting states of different parities is an equal weight mixed particle-hole mode. In correspondence with~\cite{sajith2024signatures}, we anticipate $\alpha_j = \alpha_{L+1 - j}$ and $ \beta_j = - \beta_{L+1 -j}$, which are significant only near the edges. 
Note that we interpret the `hole' contribution in this number-conserving setting as a fermionic annihilation operator acting on top of Cooper pair creation,  $C^\dag=\sum_N\ket{N+ 2}\bra{N}$, on a state $\ket{N}_e$. 

Imposing normalization on the states in Eqs.~(\ref{eq:ansE}) yields
\begin{align}
    2 &= \left( \alpha_j \alpha_k - \beta_j \beta_k \right) \Delta M^{jk} + 2 \beta_j \alpha_k \Delta P^{jk},  \label{eq:DM} \\
    0 &= \left( \beta_j \beta_k - \alpha_j \alpha_k \right) M^{jk} + 2 \beta_j \alpha_k P^{jk}, \label{eq:DM2}
\end{align}
where we introduced the anomalous correlation matrix $P^{jk}_{e/o} = \langle N+2 | c^\dagger_j c^\dagger_k | N\rangle$ for $N$ even/odd, and defined $M \equiv (M_e + M_o)/2$ and $\Delta M \equiv M_o - M_e$ (and analogously for $P$) in anticipation that $\Delta M$ and $\Delta P$ isolate the inter-edge correlations associated with the topological phase in the large $N$ limit. (Einstein summation over repeated indices is implied.)
Given the observed low-rank of $\Delta M$ and the eigenvalues $\pm 1/2$, it is natural to consider the ansatz $\Delta M^{jk} = (\beta_j \beta_k - \alpha_j \alpha_k) \tilde{M}$ with $\tilde{M} 
\approx 1/2$. Eq.~(\ref{eq:DM}) is then satisfied by choosing $\Delta P^{jk} = ( \beta_j \alpha_k - \beta_k \alpha_j) \tilde{P}$, with $\tilde{P} = 1/2$. For number-projected Kitaev wavefunctions, this symmetry between regular and anomalous correlators is indeed present~\cite{sajith2024signatures}.
More generally, there is a physical principle why we can expect $\tilde{M} = \tilde{P}$, and thus $\tilde{M} = 1/2$, in agreement with the numerically obtained eigenvalues of $\Delta M$. A non-vanishing nonlocal correlator $\Delta M^{1,L} = \bra{N}{c^\dagger_1 c_L}\ket{N}$ seems to allow instantaneous teleportation of a particle from one edge of the wire to another, while remaining in the ground state \cite{semenoff2006teleportation, fu2010electron}. To avoid this, it is necessary that the state $c^\dagger_1 \ket{N}$ have equal overlap with both $c^\dagger_L \ket{N}$ and $c_L \ket{N+2}$, which requires that inter-edge correlators in $M_{ij}$ and $P_{ij}$ (represented by $\Delta M$ and $\Delta P$) are identical in amplitude, giving $\tilde{M} = \tilde{P}$.

\emph{MZM operators.} The above analysis points to the existence of a fermionic mode whose occupation number switches between  even- and odd-parity states. The creation operator associated with this mode is given by
\begin{align}
    d^\dagger &= \frac{1}{\sqrt{2}} \sum_j \alpha_j c^\dagger_j + \beta_j c_j C^\dagger, \label{eq:d}
\end{align}
which acts as $\ket{N+1}_o = d^\dag \ket{N}_e$ and $\ket{N}_e = d \ket{N+1}_o$, as can be seen from (\ref{eq:ansE}). 
The first term preferentially creates extra density at the edges, while the second term first adds an extra Cooper pair, leading to a $2/L$ change in density uniformly throughout the system, and then removes density near the edges. Given the equal amplitude of particle and hole contributions at the edges, the net effect is an extra particle smeared uniformly through the system. Thus, even though $d$ nominally looks like a local operator, its effect is global. Moreover, the change in density $\sim 1/L$ vanishes in the thermodynamic limit and thus does not lead to locally observable consequences, as should be the case in the topological phase. 

We can now relate the operator $d$ to the enigmatic MZM operators. Due to the bimodal support of $d$ and the special relationship between the vectors $\alpha$ and $\beta$, the operators
$\Gamma^\dag_L = d^\dag + d C^\dag= 1/\sqrt2\sum_j(\alpha_j - \beta_j)(c_j^\dag + c_j C^\dag)$ and $\Gamma^\dag_R = i (d^\dag - d C^\dag) = i/\sqrt2\sum_j(\alpha_j + \beta_j)(c_j^\dag - c_j C^\dag)$ appear to be localized to a single edge each. Thus, the combinations $\alpha_j \pm \beta_j$ can be thought of as MZM wave functions (shown as green and red lines in Fig. \ref{fig:spectra}d). 
However, just as in the case of the operator $d$, the effect of $\Gamma_{L,R}$ is global due to the presence of the Cooper pair operator, increasing the density in the system uniformly by $1/L$.  

We finally note that the canonical self-conjugate Majorana operators $\gamma_L = 1/\sqrt2\sum_j(\alpha_j - \beta_j)(c_j^\dag + c_j)$, $\gamma_R = i/\sqrt2\sum_j(\alpha_j + \beta_j)(c_j^\dag - c_j)$ retain significance beyond mean-field theory. The eigenvalues and real-symmetry of $\Delta M$ imply that $\bra{N} i\gamma_R\gamma_L\ket{N}_e - \bra{N} i\gamma_R\gamma_L\ket{N}_o \approx 1$ when evaluated over fixed number states, where the subscript $o$ denotes the average expectation value over the odd parity ground states $\ket{N\pm 1}$. If $|\bra{N} i\gamma_R\gamma_L\ket{N}_e| \neq |\bra{N} i\gamma_R\gamma_L\ket{N}_o|$, the even- and odd-parity ground states would be locally distinguishable, immediately implying that $\bra{N} i\gamma_R\gamma_L\ket{N}\approx 0.5 \times (-1)^N$ in accordance with numerical results (e.g., $\sim 99\%$ accuracy for $L=64$).

\emph{Conclusion and outlook.}
In this work, we have investigated the manifestations of MZMs in a number-conserving setting. While we focused on the interacting Richardson-Gaudin-Kitaev chain, our results are more general, as we explicitly demonstrate in the End Matter. We found that the topological phase, unlike the trivial phase, is highly sensitive to the boundary conditions: the sign of the odd-even parity energy gap changes between periodic and antiperiodic boundary conditions, and vanishes with increasing system size for open boundary conditions. Even more remarkably, the single-particle correlation matrix acquires non-local parity-dependent inter-edge contributions that remain large as the system size increases. From the correlation matrix, we reconstructed the non-local fermionic mode that
converts between different number states and can be related to the Majorana modes. 
The sensitivity to boundary conditions and the existence of nonlocal boundary modes is most likely related to the Cooper pair wavefunction becoming long-ranged in topological superconductors \cite{read2000paired}. Both this exotic behavior of the pair wavefunction and the MZM modes may be accessible by measuring the correlation matrix in cold atom gas experiments. We leave detailed study of the pair wavefunction, more general interactions cases, and braiding for future work.

\emph{Acknowledgments:}
We acknowledge discussion with S. K. Yip and E.-A. Kim. JTM would  like to thank the University of Chicago REU program. IM and KA  acknowledge support by the US Department of Energy, Office of Science, Basic Energy Sciences, Materials Sciences and Engineering Division. 

\nocite{SM}
\nocite{mallik2022superfluid}
\nocite{hofmann2022heuristic}
\bibliographystyle{apsrev4-2}
\bibliography{majorana}

\section{End matter}

\begin{figure}
\includegraphics[scale=0.4]{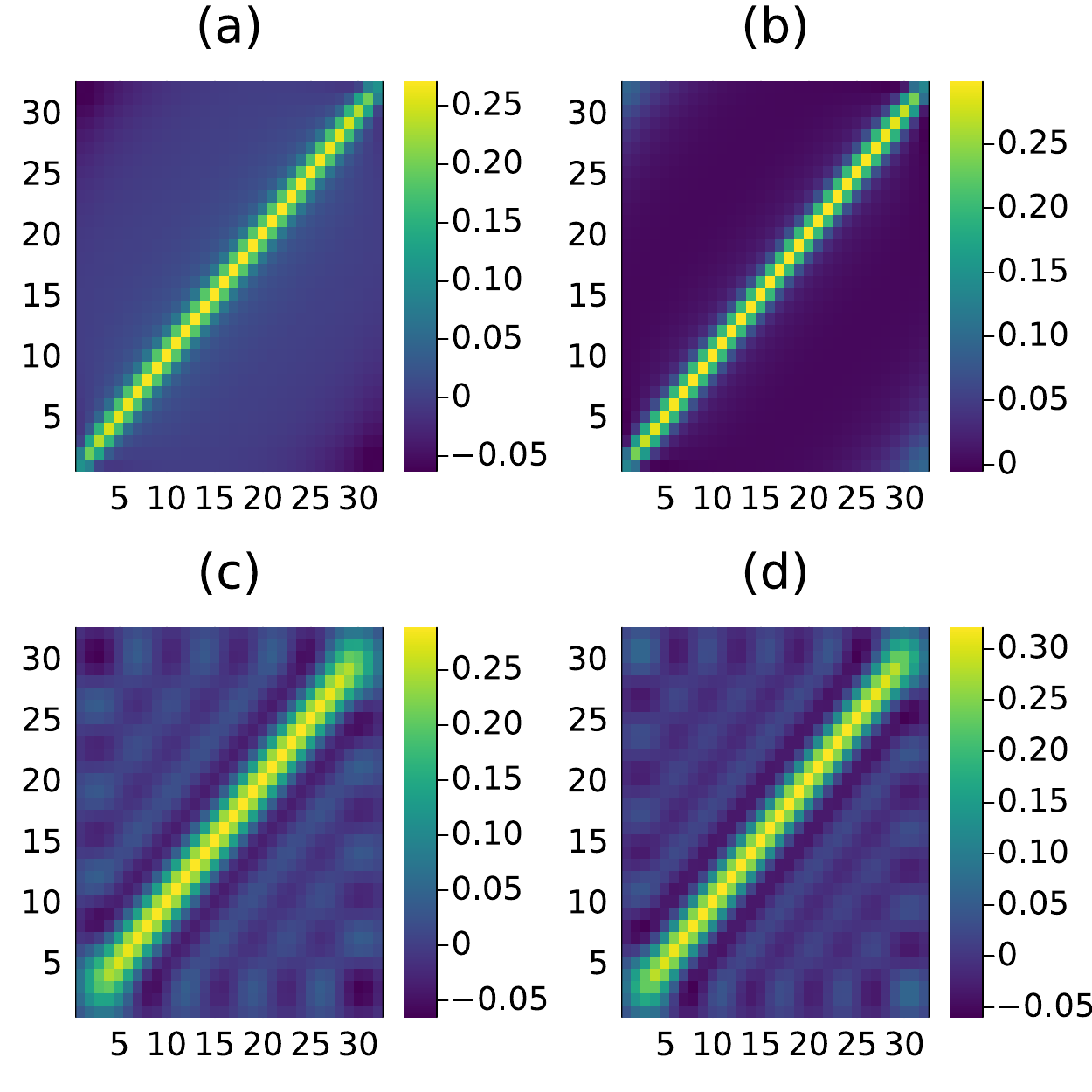}
\caption{Heat maps of the two point correlation matrix $\langle c_i^\dag c_j\rangle$ for the model defined by (\ref{eq:jaden}, \ref{eq:small}) with  $g=1$, $L=32$ for (a) $N=8$ and (b) $N=9$; and the model defined by (\ref{eq:NN}) with $g=0.4$, $L=32$ for (c) $N=8$ and (d) $N=9$. In the gapless model (\ref{eq:NN}), the MZM correlations exhibit stronger oscillatory (checkerboard) behavior due to the higher susceptibility of the bulk low-energy modes.
}
\label{fig:corrs_NN}
\end{figure}
\begin{figure}
\includegraphics[scale=0.4]{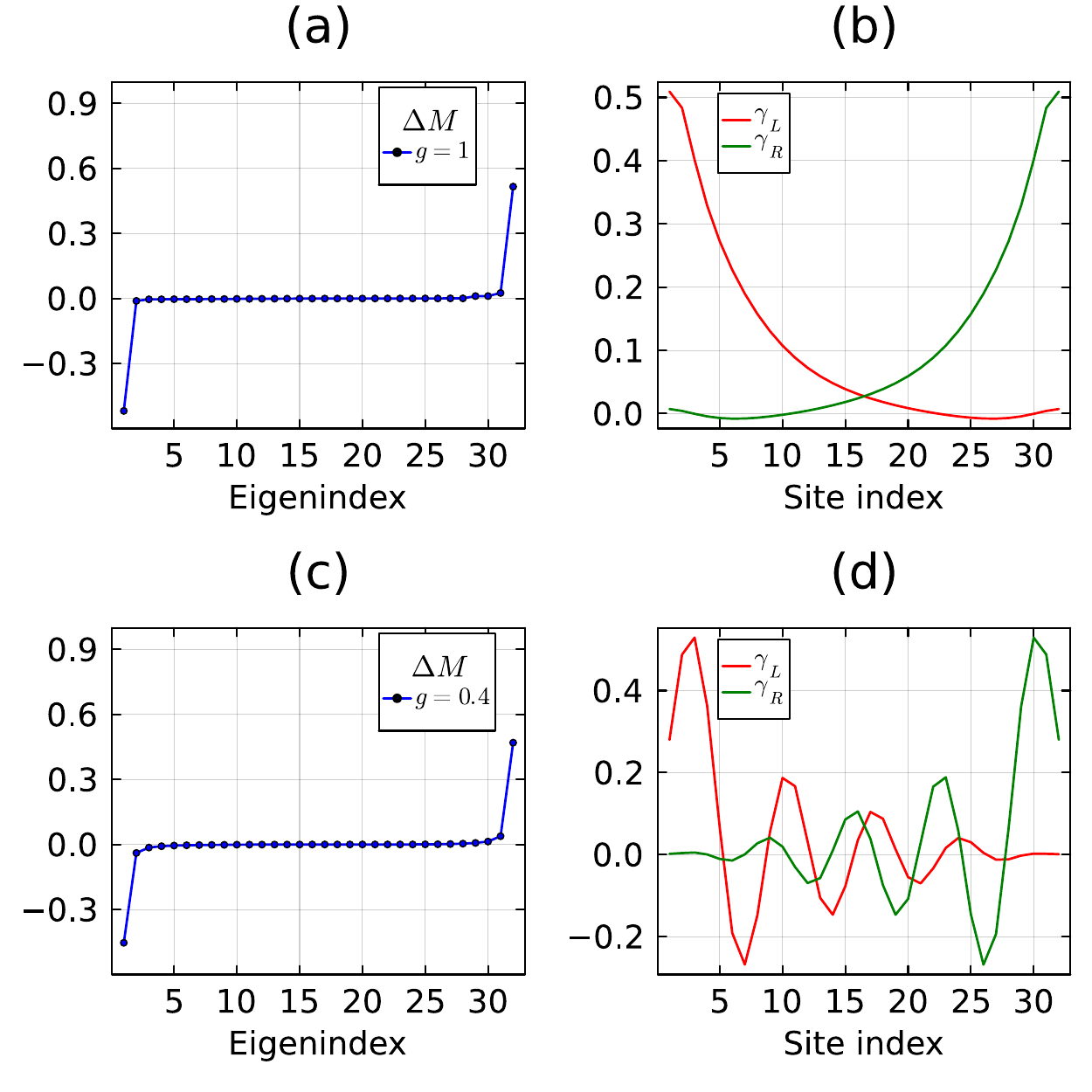}
\caption{Spectra of correlation matrix (a) $\Delta M$ for the model defined by (\ref{eq:jaden}, \ref{eq:small}) with $g=1$ and $L=32$ at $1/4$ filling and (b) the sum and difference of the eigenvectors of $\Delta M$ with extremal eigenvalues; (c) $\Delta M$ for the model defined by (\ref{eq:NN}) with $g=0.4$ and $L=32$ at $1/4$ filling and (d) the sum and difference of the eigenvectors of $\Delta M$ with extremal eigenvalues.}
\label{fig:spectra_NN}
\end{figure}

The algebraic decay of the RGK Majorana wavefunctions raises the question of whether the associated modes can be regarded as strictly zero-energy, since their power-law tails may generate non-vanishing overlap contributions even in the thermodynamic limit. We emphasize, however, that this feature is an artifact of the specific form of the RGK interaction. Although our analysis has focused on the RGK model, the underlying results are more general: the eigenvectors of $\Delta M$ provide direct access to the Majorana wavefunctions for a broad class of number-conserving Hamiltonians supporting such boundary modes. This includes models obtained by modifying the pairing potential $\eta (r)$, or even those with purely density-density interactions. A particularly simple example is
\begin{align}
    \mathcal H = -\sum_{i=1}^{L-1} (c_i^{\dagger}c_{i+1} + h.c.) - \frac{4g}{L} I_{\phi}^{\dagger} I_{\phi}, \label{eq:jaden}
\end{align}
with $I_\phi$ as defined in (3) and
\begin{align}
    \eta (r) = \begin{cases} 
           1 & r=1 \\
          0 & r>1
       \end{cases}.\label{eq:small}
\end{align}
Despite appearances, the interaction is  non-local, corresponding to local (nearest-neighbor) Cooper pairs hopping arbitrary distance with the same matrix element.
At the  mean-field level, this model reduces to the standard Kitaev chain. In Fig.~\ref{fig:corrs_NN}, we show heatmaps of the two-point correlation matrix $\langle c_i^\dag c_j\rangle$ for $g=1$, $L=32$, computed for (a) $N=8$ and (b) $N=9$, and in Fig.~\ref{fig:spectra_NN} we show (a) the spectrum of $\Delta M$ and (b) the corresponding MZM wave functions for $g=1$ at $1/4$ filling. As in the RGK case, to a very good approximation, $\Delta M$ is rank-2 with eigenvalues $\pm1/2$, and the resulting edge modes exhibit exponential decay consistent with Kitaev’s mean-field solution. This  contrasts with the algebraically localized MZMs revealed by $\Delta M$ in the RGK case. We also note that the bulk spectrum of this model is gapped (as is the RGK spectrum), which is beneficial for isolating the Majorana sector during adiabatic manipulations such as braiding. These features make the model (\ref{eq:jaden},\ref{eq:small}) a promising candidate for further theoretical exploration. 

For completeness, we  also consider a Hamiltonian with truly short range density-density interactions:
\begin{align}
    \mathcal H = -\sum_{i=1}^{L-1} (c_i^{\dagger}c_{i+1} + h.c.) - 4g \sum_{i=1}^{L-1} n_i n_{i+1}. \label{eq:NN}
\end{align}
Since the interactions are strictly local, one does not expect bulk long range order or a spectral gap.
In Fig.~\ref{fig:spectra_NN}, we present heatmaps of the two-point correlation matrix $\langle c_i^\dag c_j\rangle$ for $g=0.4$, $L=32$ and (c) $N=8$ and (d) $N=9$. The eigenvectors of $\Delta M$ corresponding to the $\pm 0.5$ eigenvalues again exhibit edge-localized profiles, while $\Delta M$ remains rank-2 to a very good approximation. In Fig.~\ref{fig:spectra_NN}, we show (c) the spectrum of $\Delta M$ and (d) the resulting MZM wave functions. The edge modes exhibit oscillatory decay, with a localization length that decreases as $g$ is increased (up to the phase boundary). Near $g=0.5$ the system undergoes a transition into a phase-separated state that breaks reflection symmetry.
\setcounter{secnumdepth}{2}

\section*{Supplemental Material}

\section{Spectra of closed RGK Chain}\label{sup:sec:closed}

\begin{figure}
\includegraphics[scale=0.40]{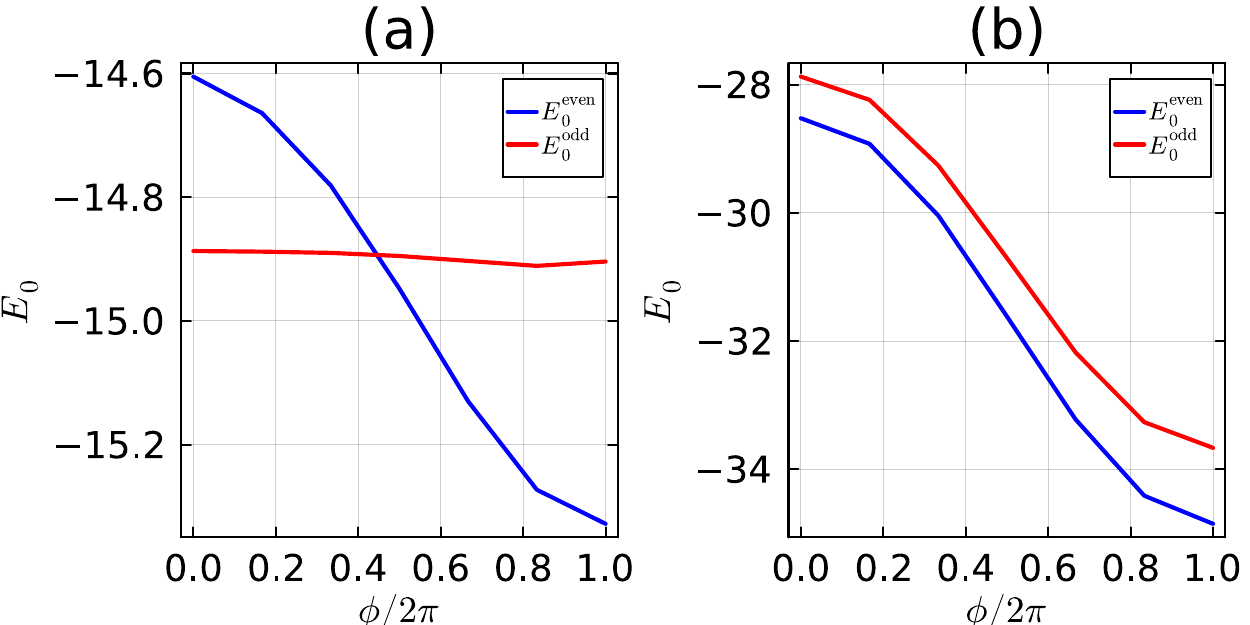}
\caption{Even- and odd-parity ground state energies as a function of the boundary condition phase twist, $\phi$, for $L=32$ and $N=8$. $\phi=0$ ($\phi=2\pi$) corresponds to periodic (antiperiodic) boundary conditions. Here, $E_0^{\textnormal{even}}(N,L) = E_0(N,L)$ and $E_0^{\textnormal{odd}}(N,L) = (E_0(N+1,L) + E_0(N-1,L))/2$. Plot (a) is for $g=1$ (topological) and plot (b) is for $g=3$ (trivial). 
}\label{fig:E_gap1}
\end{figure}

Consider slowly changing the boundary conditions from periodic to antiperiodic. A topological invariant $Q$ can be defined as the difference in effective ground state fermion parities between chains with periodic and antiperiodic boundary conditions~\cite{ortiz2014many}. In particular, an odd number of gap closings corresponds to a nontrivial $Q$, signaling that the superconductor is topologically nontrivial. We illustrate this behavior in Fig. \ref{fig:E_gap1}. 

\begin{figure}
\includegraphics[scale=0.40]{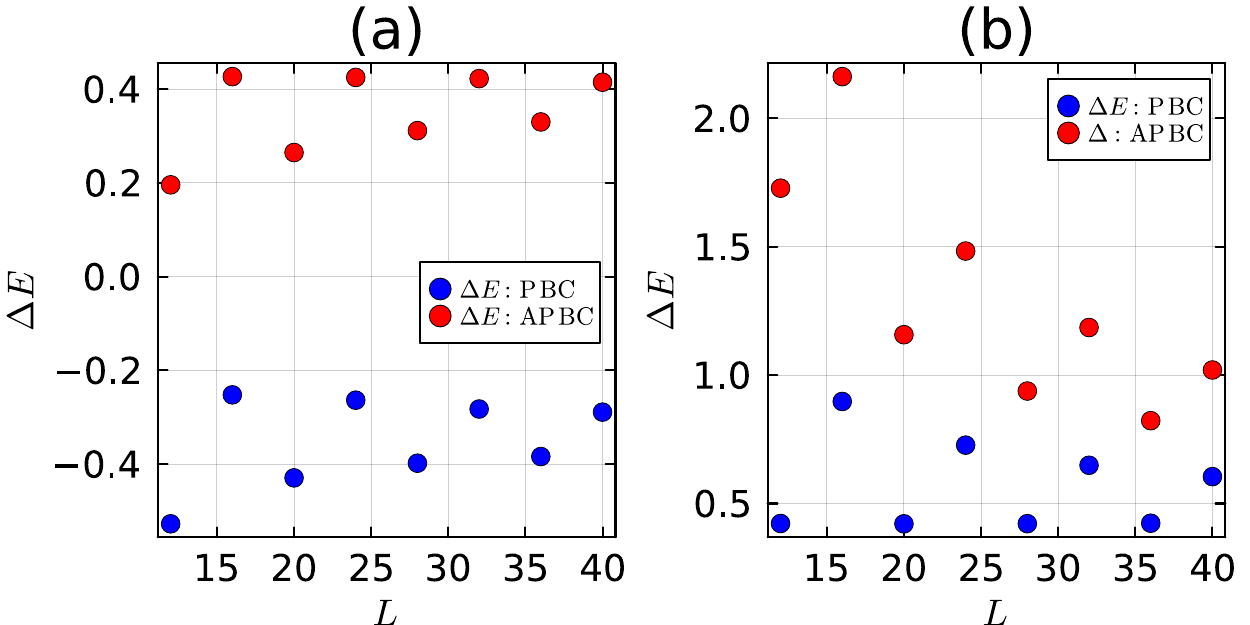}
\caption{Odd-even excitation gap vs system size $L$ ($N=L/4$) for periodic and antiperiodic boundary conditions, for (a) $g=1$ (topological) and (b) $g=3$ (trivial).}
\label{fig:E_gap_closed}
\end{figure}

Fig. \ref{fig:E_gap_closed} shows the odd-even excitation gap $\Delta E$ (defined in the main text) as a function of system size $L$ for periodic and antiperiodic boundary conditions, for $g=1$ and $g=3$. 
For $g=1$, in the topological phase, $\Delta E$ does not decay to zero.  
Under periodic boundary conditions, the system favors the odd-parity ground state, regardless of the parity of $N$, even though there are some oscillations as a function of $N$, with the odd-parity state being somewhat more favored for odd $N$  relative to even $N$.  For antiperiodic boundary conditions, however, the favored ground state flips---the system now prefers the even-parity ground state. This boundary-condition–sensitive parity inversion can be viewed as a signature of nontrivial topology. 

On the other hand, in the trivial phase realized for larger $g$, the even-parity sector is favored for any $N$ and any  boundary conditions. This aligns with conventional superconducting behavior and highlights the distinction between the phases. 

\section{Evidence for BEC in trivial phase}

\begin{figure}
\includegraphics[scale=0.40]{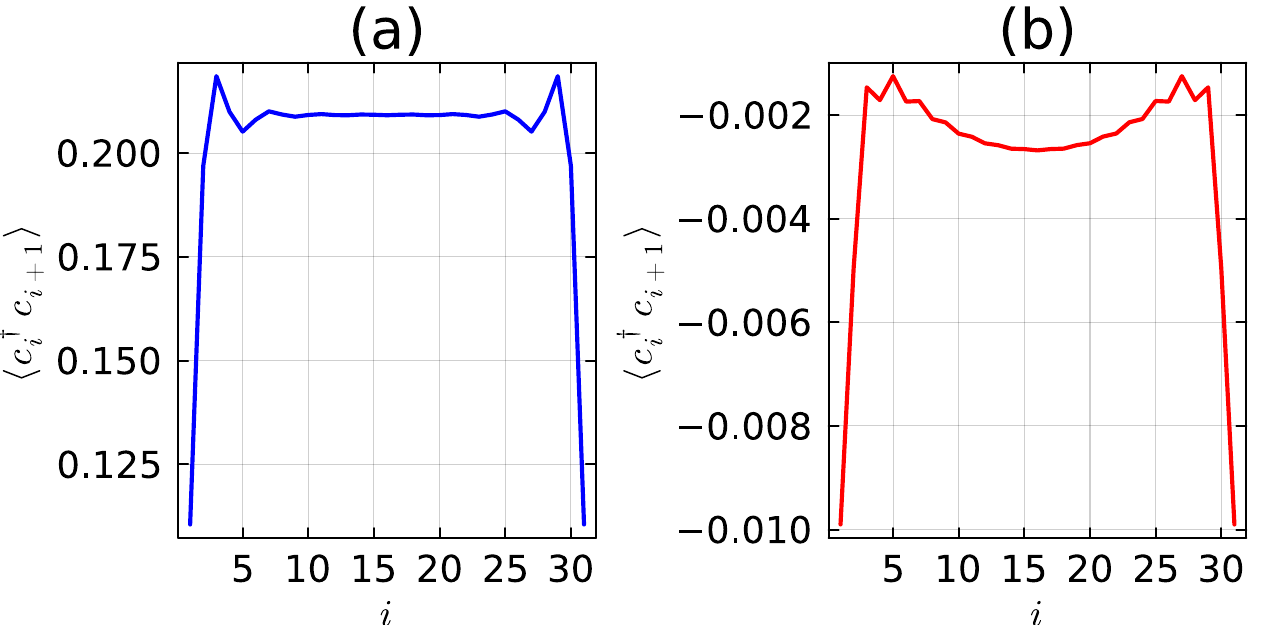}
\caption{(a) $\langle c_{i}^\dagger c_{i+1}\rangle$ for $L=32, N=8$ and $g=1$. (b) $\langle c_{i}^\dagger c_{i+1}\rangle$ for $L=32, N=8$ and $g=3$. $\langle c_{i+1}^\dagger c_i\rangle$ is two orders of magnitude larger for $g=1$ (topological) than for $g=3$ (trivial). 
} 
\label{fig:densitites1}
\end{figure}

In Fig.~\ref{fig:densitites1}, we observe a dramatic change in $\langle c_i^\dagger c_{i+1}\rangle$ across the phase boundary: the correlator drops by two orders of magnitude between $g = 1 $ and $g=3.$ This correlator measures kinetic energy. 

There are two possible interpretations of this effect: the formation of small ``heavy" Cooper pairs or macroscopic phase separation. In the former case, the state would remain a (BEC) superconductor, while in the latter it would be an insulator. We note that the model with only short-range attraction exhibits phase separation at strong coupling (the Ising regime of the corresponding XXZ spin model~\cite{sajith2024signatures}). However, in long-range interacting RGK models, where interaction alternates in space between attraction and repulsion, the phase separation can be averted in favor of BEC. Indeed, the preference for even fermion parity states that we found both for closed and open  boundary conditions for $g = 3$ support the Cooper pair BEC scenario. So does the significant superfluid stiffness, which manifests in the energy dependence on the boundary conditions under phase twist (Fig \ref{fig:E_gap1}b)~\cite{Mallik2022}.  The kinetic energy in this case is determined by the hopping of Cooper pairs, which for the nearest-neighbor Cooper pairs can be estimated as $t^2/U$, where $U\sim g$ is the nearest-neighbor attraction ~\cite{Hofmann2022}.

\section{Number-projected BCS ansatz}\label{sec:proj} The limited change in the spectrum of the correlation matrix $M$ between different parity states in the trivial phase hints at an underlying mean-field description, whereby the change in the correlation matrix occurs due to the occupation/emptying of a single mode. Following BCS, we propose even and odd particle number wavefunctions 
\begin{align}
\ket{N}_e &\propto P_N \cdot e^{A_{ij} c^\dagger_i c^\dagger_j} \ket{0} = P_N \cdot \prod_l e^{\lambda_l \tilde{c}^\dagger_{l,1} \tilde{c}^\dagger_{l,2}} \ket{0}, \nonumber \\
\ket{N+1}_o &\propto \tilde{c}_{\ell,1}^\dagger \ket{N}_e . \label{eq:BCSodd}
\end{align}
Here, $\sum_{ij} A_{ij} c^\dagger_i c^\dagger_j$ is the Cooper-pair creation operator (with $A_{ij}$ a skew-symmetric matrix), and the projection operator $P_N$ for $N$ even simply selects the state with $N/2$ Cooper pairs. This wavefunction is a number projected mean-field BCS state. The operators $\tilde{c}_{l,1}, \tilde{c}_{l,2}$ represent orthonormal modes in the canonical basis of $A_{ij}$, with eigenvalues $\lambda_l > 0$.  The odd parity state can be obtained by breaking one, say the $\ell^{\text{th}}$ `component' of the Cooper pair wavefunction and fixing the occupation probability of the modes $\tilde{c}^\dagger_{l = \ell, 1 }$ and $\tilde{c}^\dagger_{l = \ell, 2 }$ to $1$ and $0$, respectively. 
We note here that on a ring with periodic boundary conditions there are two special modes, $k = 0$ and $k=\pi$, which lack time-reversed partners, and should be treated separately \cite{ortiz2014many, wang2017number}.

The wavefunction $\ket{N}_e$ yields a correlation matrix $M$ with a fully doubly degenerate spectrum and eigenvalues $\lambda_l^2/ (1 +  \lambda_l^2)$ (in the large $N, L$ limit). The derivation is trivial for the non-projected BCS wavefunction; then, since that the operator $c^\dag_i c_j$ does not change the number sector, the result must match the most probable number projected states (ones with $N \approx \langle N\rangle $), up to $\mathcal{O} \left( 1/N \right)$ corrections to the eigenvalues. 

In the odd parity state, the spectrum should remain nearly identical, except for the doubly-degenerate eigenvalues associated with the $\ell^{\text{th}}$ pair of states being replaced by eigenvalues $0$ and $1$. $\Delta M$ is then a low-rank matrix with non-zero eigenvalues $- \lambda_\ell^2 / (1 + \lambda_\ell^2), 1/(1 + \lambda_\ell^2) $. Within this approximation, the orthonormal modes are identical for $M_e, M_o$ and $\Delta M$. 

We find that the above ansatz works well in the trivial phase for both open and periodic boundary conditions.  In the odd parity state, the mode with $k \approx 0$, which was empty in the even $N$ state, becomes occupied (appearing with an eigenvalue $1$ in the correlation spectrum). 

\begin{figure} 
\includegraphics[scale=0.40]{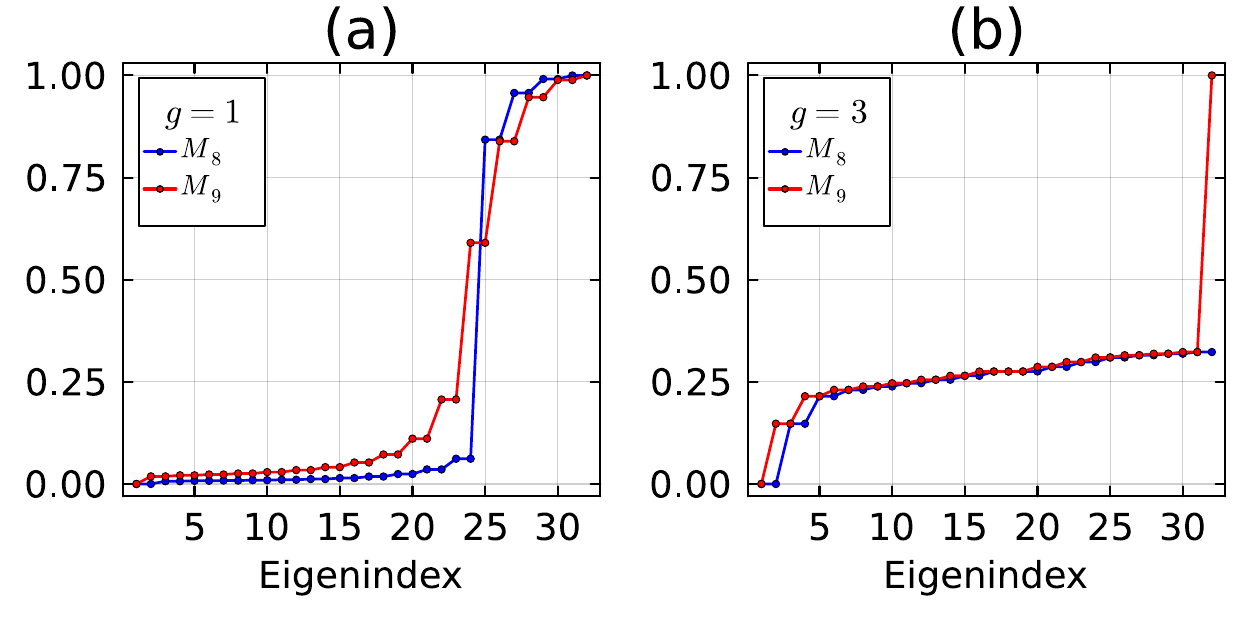} 
\caption{The correlation spectrum for the even and odd states at $N = 8,9$ for $L = 32$ with periodic boundary conditions in (a) the topological phase ($g = 1$) and (b) the trivial phase ($g = 3$). In the topological phase, both odd and even states have the modes $k = 0$ $(\pi)$ occupied (empty). The even state corresponds to the odd state but with one  pair of modes with $\lambda_\ell \approx 1$ (eigenvalue $1/2$ in the correlation matrix) broken up into fully occupied and fully empty modes. In the trivial phase, the even parity state has all modes paired up, while in the odd parity state the $k = 0$ mode switches from being empty to fully occupied.
}
\label{fig:topopbc}
\end{figure}

The spectrum of $ M$  for the topological phase with periodic boundary conditions can be explained equally well with this ansatz, with the modification that the $k=0$ mode remains unpaired and occupied in both the odd and even parity states. 
That makes the odd-parity state of the form
\begin{equation}
\ket{N+1}_o \propto c^\dag_{k=0}P_N \cdot \prod_k e^{\lambda_k \tilde{c}^\dagger_{k} \tilde{c}^\dagger_{-k}} \ket{0} 
\end{equation}
a good reference  in this case.
The Cooper pair operator includes all momenta except $0$ and $\pi$, which do not have time-reversed partners.
This odd-parity ground state does not have any excited quasiparticles.
Instead, it is the even-parity state that has one Cooper pair unbound, with the wavefunction
\begin{equation}
    \ket{N}_e \propto \tilde{c}_{\ell,1}^\dagger \ket{N-1}_o. 
\end{equation}
In Fig.~\ref{fig:topopbc}, we show how the spectrum changes for $L = 32$ between $N = 8$ and $9$ with periodic boundary conditions. In the odd parity state ($N = 9$), there are two distinct modes at $n = 0$ and $1$, which upon inspection are found to be the $k = \pi$ and $0$ modes, respectively. Also, this state has a pair of modes near $n \approx 0.5$. These are a pair of modes with intermediate values of momenta, $(k_*, -k_*)$. They seem to vanish in the $N = 8$ state. However, in reality, they split up, ending near $n = 0$ and 1, as expected from our ansatz for the topological state with periodic boundary conditions. 

In contrast, the topological state with open boundary conditions, which is the focus of the main text, presents a significant challenge.
As can be seen from Fig. 3a in the main text, the difference between odd and even $N$ states is subtle, with the spectra changing gradually with $N$. Moreover, the eigenmodes of $\Delta M$ with eigenvalues $\pm 1/2$ are not eigenstates of $M_e, M_o$, which is also inconsistent with the simple projected BCS ansatz.
For certain, there is no pair unbinding near the Fermi level (near occupancy $n\sim 0.5$). The notable difference between the odd and even cases is that in the odd-number state, there are two distinct unpaired modes with occupancies very close to $0$ and $1$.  
We discuss the nature of these modes in the following section. It turns out that the latter (fully occupied) mode is a direct descendant of the $k = 0$ mode in a closed system, while the empty mode descends from the $k = \pi $ mode (see Fig.~\ref{fig:special_mode}).

\paragraph{Zak phase.---} The span of the eigenvalues $\lambda_l$ differs in the two phases: in the topological phase, the eigenvalues $n_l = \lambda^2_l/(\lambda^2_l + 1)$  span the entire range from $0$ to $1$ smoothly in both even and odd parity states. This is  in correspondence with mean-field expectations. Under periodic boundary conditions, the eigenmodes are  pairs with momenta $(k, -k)$. The eigenvalues associated with these modes $\lambda_k$ are $v_k/u_k$, where the coefficients $v_k, u_k$ are the usual Bogoluibov coefficients that appear in the BCS wavefunction $\psi \sim \prod \limits_{k} \left( u_k + v_k c^\dagger_{k} c^\dagger_{-k} \right) \ket{0}$. For example, from the mean-field solution of Kitaev model, $\lambda_k = \frac{v_k}{u_k} = \frac{E_k - \xi_k}{E_k + \xi_k} = \frac{\Delta_k^2}{(E_k + \xi_k)^2}$, where $\xi_k = -2t \cos k - \mu, \Delta_k = \Delta \sin k$ and $E_k = \sqrt{\Delta^2_k + \xi^2_k}$. The topological phase is obtained for $-2t < \mu < 2t $, that is, while the chemical potential lies inside the band. Also, $\Delta_k = 0$ at the edges of the band (at $k = 0,$ $\pi$). This implies $E_k = \abs{\xi_k}$ at the edges of the band as well. Clearly, in the trivial phase, the demoninator $E_k + \xi_k$ is always finite for $\mu < -2t$, as $\xi_k > 0$ for all $k\in(0,\pi)$. Thus, $\lambda_k$ is always finite and the range of eigenvalues in the correlation matrix does not approach $1$ (in the even $N$ state). In the topological phase, the denominator must vanish at one edge of the band (and faster than the numerator), while the numerator vanishes at the other edge. This implies that the eigenvalues $\abs{\lambda_k}$ span $(0, \infty)$ and the eigenvalues of the correlation matrix span $(0,1)$. 

The full span of the correlation matrix spectrum correlates with the non-trivial winding of the Zak phase of the single-particle wavefunction $\tan^{-1} \left( \Delta_k/ \xi_k \right)$. The
Zak phase winds from $0$ to $2\pi$ as momentum runs from $k = -\pi$ to $ k = \pi$ in the topological phase and does not wind in the trivial phase.

\section{The special eigenvalue-1  mode in the topological phase with odd $N$}\label{sec:ev1}
\begin{figure}
\includegraphics[scale=0.40]{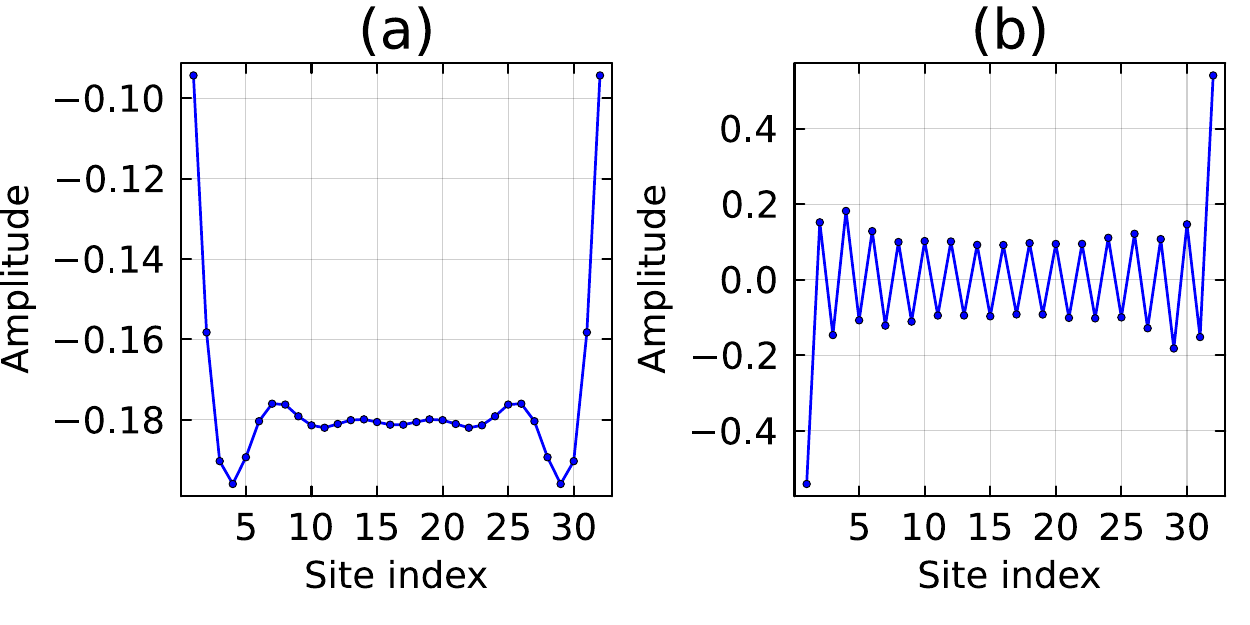}
\caption{(a) Eigenvalue-$1$ and (b) eigenvalue-$0$ mode (eigenvector) of the correlation matrix $M_9$ for $L=32$ and $g = 1$.}
\label{fig:special_mode}
\end{figure}

Even though the simple projected BCS ansatz does not fully capture the odd-parity state in the topological phase, a careful examination of the eigenstates of $M_o$ reveals the appearance of an unpaired fully populated fermion mode. The spatial structure of this mode is similar to the fermion mode in \cite{wang2017number}, which converts between even and odd fermion number Kitaev ground states, $\ket{\Psi^{N+1}_o} \propto a_1^\dag \ket{\Psi_e^{N}}$. They found that $a_1$ is an equal superposition of all site modes. Indeed, this is close to what we find numerically for the eigenvector with the eigenvalue $1$ in Fig.~\ref{fig:special_mode}a.

It is easy to prove that if $\ket{\Psi^{N+1}_o} \propto a_1^\dag \ket{\Psi_e^{N}}$, there must be an eigenvalue $1$ in the correlation matrix.
First let's normalize, $\ket{\Psi^{N+1}_o} = a_1^\dag \ket{\Psi_e^{N}}/\sqrt{C}$, where $C = \bra{\Psi_e^{N}}a_1 a_1^\dag |\Psi_e^{N}\rangle$. Now,  expand site fermions $c_j$ in terms of orthogonal modes $a_1, \  a_2, ..., a_{L}$, where all modes except $a_1$ are chosen arbitrarily, $c_j = A_{jn} a_n$, and $A$ is an orthogonal matrix. Then 
\begin{eqnarray}
    M^o_{jk} &=& \bra{\Psi^{N+1}_o}c^\dag_jc_k \ket{\Psi^{N+1}_o} \nonumber \\
    &=& A_{jn}A_{km} \bra{\Psi^{N+1}_o}a^\dag_n a_m \ket{\Psi^{N+1}_o} \nonumber \\
    &=& A_{jn}A_{km} \bra{\Psi^{N}_e}a_1 a^\dag_n a_m a_1^\dag\ket{\Psi^{N}_e}/C \nonumber \\
    &=& A_{j1}A_{k1} \bra{\Psi^{N}_e}a_1 a^\dag_1 a_1 a_1^\dag\ket{\Psi^{N}_e}/C \nonumber \\
    &+& \sum_{m,n \ne 1}^{L} A_{jn}A_{km} \bra{\Psi^{N}_e}a_1 a_1^\dag a^\dag_n a_m \ket{\Psi^{N}_e}/C \nonumber \\
    &=& A^{-1}_{1j}A^{-1}_{1k}  
    + \sum_{m,n \ne 1}^{L} A_{jn} G_{nm} A^{-1}_{mk},
\end{eqnarray}
where $G_{nm}=
\bra{\Psi^{N}_e}a_1 a_1^\dag a^\dag_n a_m \ket{\Psi^{N}_e}/C$ is a symmetric matrix of size $L-1\times L-1$. The second term therefore is rank-deficient, spanned by the vectors orthogonal to $A^{-1}_{1j}$. But $A^{-1}_{1j}$ are nothing but $v_j$ in the expansion $a_1 = v_j c_j$. Thus, we have proven that the mode $a_1^\dagger$ indeed has an associated eigenvalue $1$ in the correlation matrix.

\begin{figure}
\includegraphics[scale=0.40]{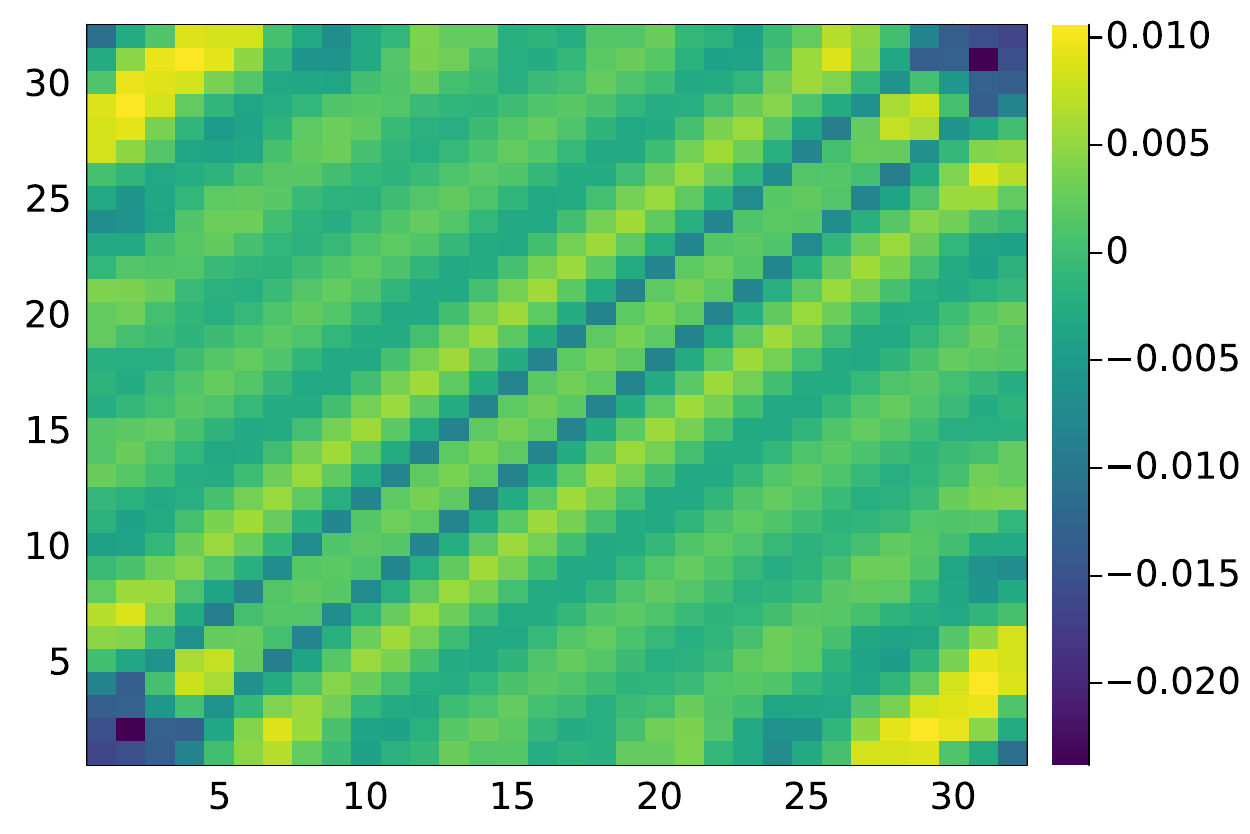}
\caption{Heatmap of the difference $\bra{N} a_1c_i^\dagger c_ja_1^\dag\ket{N}/C -\bra{N+1}c_i^\dagger c_j\ket{N+1}$ for $N=8$, $L=32$ and $g=1$. The difference is small, indicating that the operator $a_1^\dagger/\sqrt{C}$ is a good approximation of the transition between $\ket{N}_e$ and $\ket{N+1}_o$.}
\label{fig:a_1}
\end{figure}

We have verified that the mode $a_1^\dag = v_j c_j^\dag$ converts between even and odd parity states by comparing  $\bra{N}  a_1 c_i^\dag c_j a_1^\dag \ket{N}/\bra{N}  a_1 a_1^\dag \ket{N}$ with $\bra{N+1}  c_i^\dag c_j \ket{N+1}$ (see Fig. \ref{fig:a_1}). 
However, it cannot be considered a Majorana mode. It is easy to check that the weight $C$ is low: $\bra{N}  a_1 a_1^\dag \ket{N} \sim O(1/L)$, which also implies that $\bra{N}  a_1^\dag a_1 \ket{N} \sim 1- O(1/L)$.
We stress here that the ansatz of Eq. 5 in the main text conjectures an \emph{equality} that identifies a \emph{normalized} quasiparticle creation
operator in contrast to the action of $a_1$ on $\ket{N}_e$. 
We have verified numerically that for $g = 1$, $v^T M_e v \approx 0.97$. Since $v$ is normalized,  the $a_1$ mode is already  $97\%$ occupied in the even state. Acting on the even state with $a_1^\dag$ promotes the occupancy to 100\%.
The essentially uniform change in the density induced by $a^\dag_1$  is consistent with the expectation that the states with odd and even parities are locally indistinguishable in the thermodynamic limit.

\begin{figure}
\includegraphics[scale=0.40]{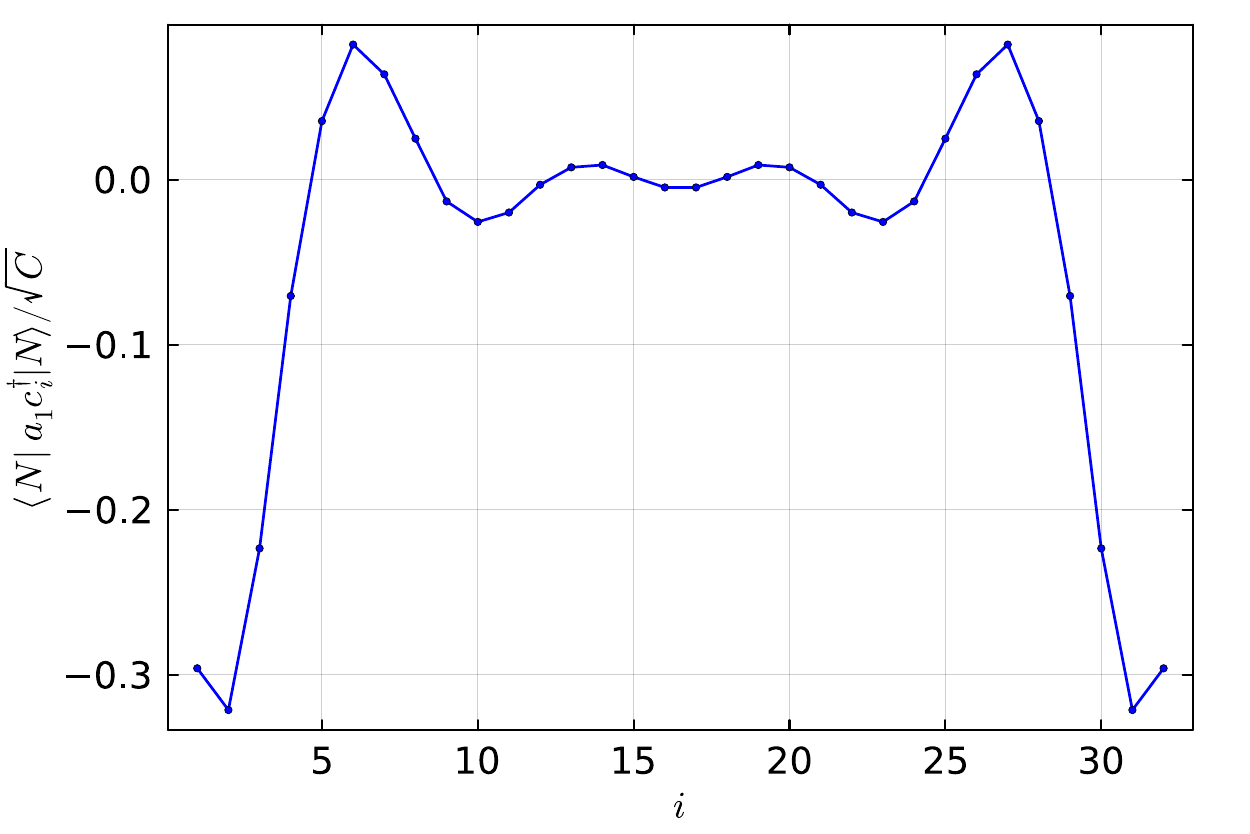}
\caption{$\langle N | a_1 c_i^\dag |N\rangle/\sqrt{C} \approx \langle N+1 | c_i^\dag |N\rangle$ for $N=8$, $L=32$ and $g=1$.} 
\label{fig:rho}
\end{figure}

The low efficiency of this operator is due to the fact that for the internal sites $i$ away from the edges, the expectation value $\bra{\Psi^{N+1}_o}c^\dag_i\ket{\Psi^{N}_e}$ vanishes rapidly, which can be seen explicitly in the example of the Kitaev wave function~\cite{sajith2024signatures}. In Fig. \ref{fig:rho}, we plot $\bra{N}a_1 c^\dag_i\ket{N}/\sqrt{C} \approx \bra{N+1}c^\dag_i\ket{N}$ which shows the same qualitative behavior.
Despite their small contribution, dropping the internal sites from the expression for $a_1^\dag$ while keeping the operator normalized would lead to exaggerated amplitudes near the edges, ``overfilling" them and causing inhomogeneous density distribution inconsistent with topological indistinguishability of odd and even  ground states. If we insist on the  operator acting ``locally" near the edges, we would need to offset the edge charge creation by charge annihilation. This is the physical origin of the hole contribution in the operator $d^\dag$ in Eq. 8 in the main text---it creates charge throughout the system and depletes it at the edges. Of course, $d^\dagger$  is not truly localized to the edges since it involves the Cooper pair creation operator that affects the density in the whole chain. 

\section{Anomalous correlators}
In Fig. \ref{fig:anomolous}, we plot the anomalous correlation matrix $P^{jk}_{e/o} = \langle N+2 | c^\dagger_j c^\dagger_k | N\rangle$ for even/odd $N$. It has a structure similar to $M^{jk}_{e/o}$, except it is antisymmetric due to Fermi exchange statistics. Notably, off-diagonal correlators are large, consistent with the no-teleportation argument (see main text). This is made clearer by Fig. \ref{fig:M_P_diff}, where we plot $\Delta M^{jk}$ and $\Delta P^{jk}$, which isolate the off-diagonal structure. We note here that there is a gauge freedom associated with these anomalous correlation matrices due to the lack of $U(1)$ symmetry. In particular, since they are real, the sign can be chosen arbitrarily to make short-range correlations independent of the parity. 

Using the anomalous correlators, we can numerically verify the normalization conditions Eqs. 6 and 7. They evaluate to $2.38$ and $-0.03$, respectively. These deviations from the expected values are not surprising, given the non-negligible difference in filling fraction between $N=8$ and $N=10$, as well as finite system size effects.

\section{MZM overlap energy}
\label{sup:sec:MZM_splitting}

\begin{figure}
\includegraphics[scale=0.40]{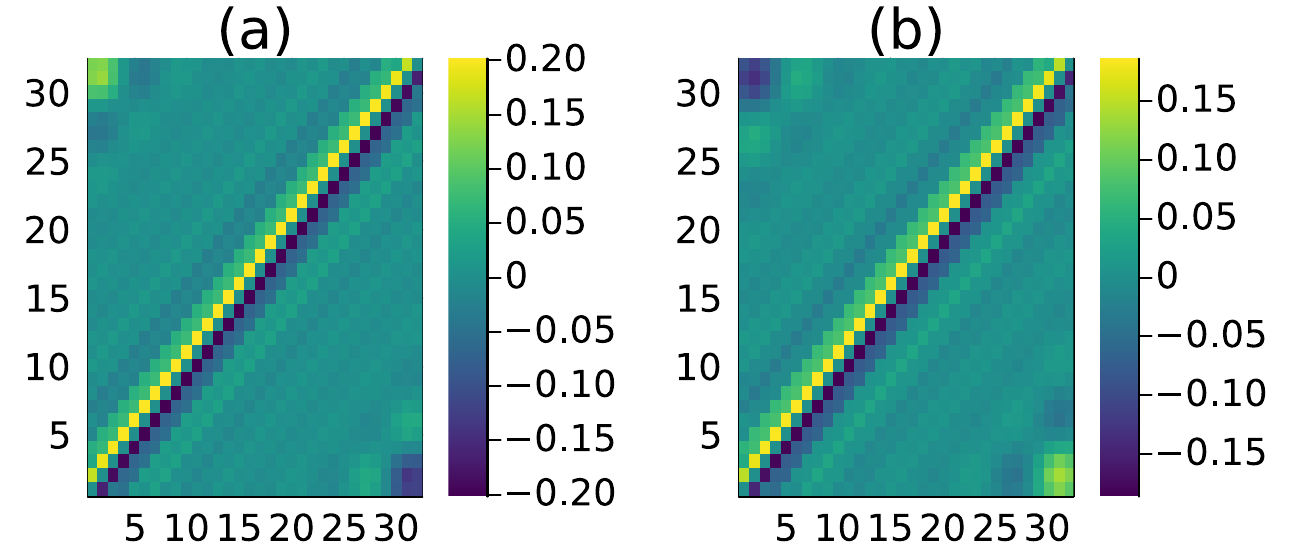}
\caption{Heat map of $P^{jk}_{e/o} = \langle N+2 | c^\dagger_j c^\dagger_k | N\rangle$ for (a) $N=8$ and (b) $N=7$, at $L=32$ and $g=1$.}
\label{fig:anomolous}
\end{figure}

\begin{figure}
\includegraphics[scale=0.40]{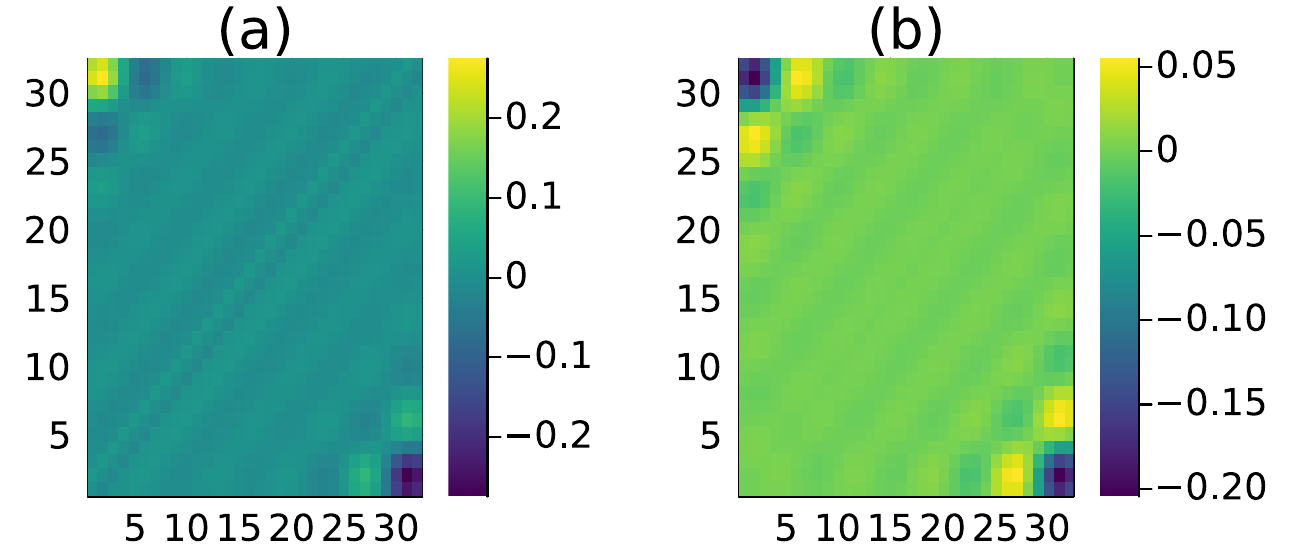}
\caption{(a) Heat map of $\Delta P^{jk} = \langle 10 | c^\dagger_j c^\dagger_k | 8\rangle - \langle 9 | c^\dagger_j c^\dagger_k | 7 \rangle$ for $L=32$ and $g=1$ (b) Heat map of $\Delta M^{jk} = \langle 8 | c^\dagger_j c_k | 8\rangle - \frac{1}{2} (\langle 7 | c^\dagger_j c_k | 7 \rangle + \langle 9 | c^\dagger_j c_k | 9 \rangle)$ for $L=32$ and $g=1$.} 
\label{fig:M_P_diff}
\end{figure} 

\begin{figure}
\includegraphics[scale=0.40]{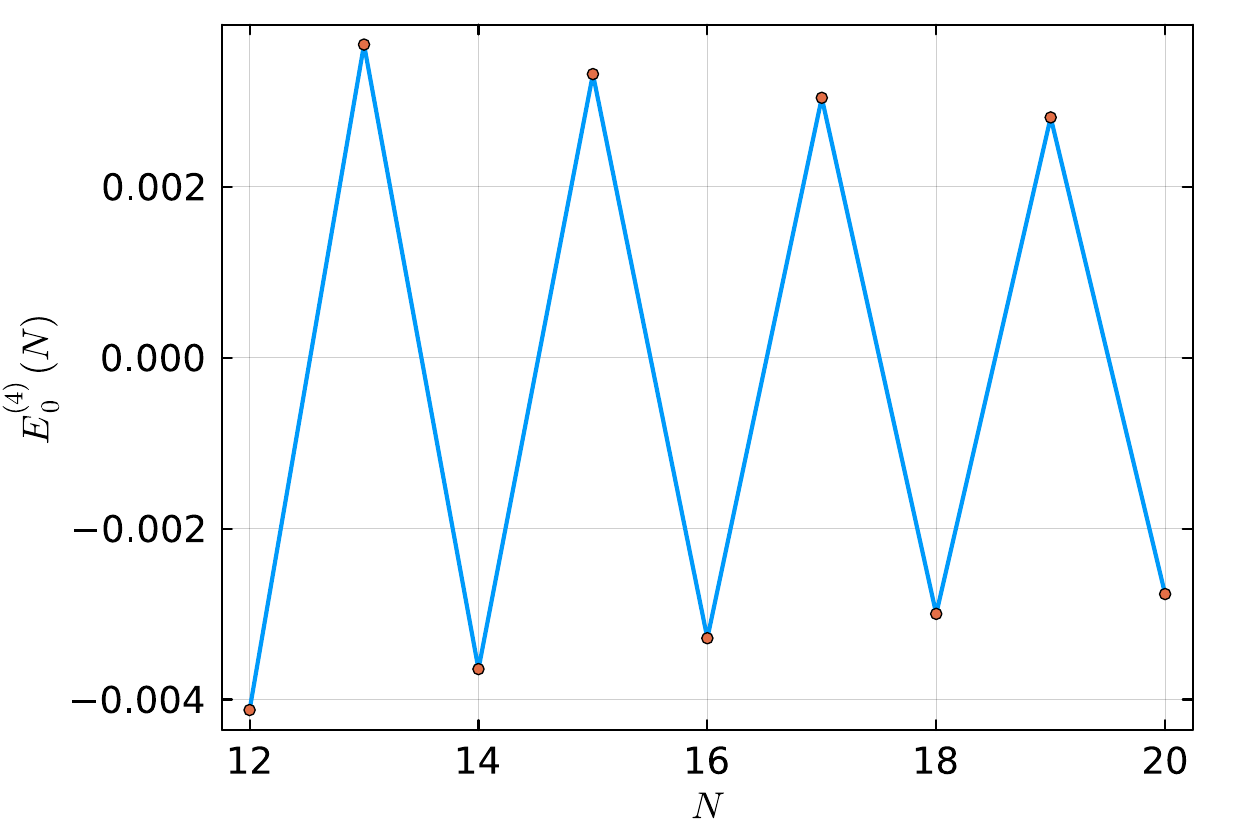}
\caption{The fourth discrete derivative of $E_0(N)$, $E^{(4)}_0(N)$, plotted for $N=12,...,20$ and $L=4N$.} 
\label{fig:E^(4)_0}
\end{figure}

The apparent smoothness of the discrete second derivative of $E_0(N)$ in Fig. 1 in the main text suggests that $E_0(N)$ itself is a smooth function of $N$. However, a more careful analysis reveals that $E_0(N)$ is in fact not smooth. There are tiny oscillations with the parity of $N$, which may be anticipated given that for finite $L$, the MZM overlap is finite, albeit small, and the occupation number of the MZM alternates between even and odd $N$. To isolate the parity-sensitive component of $E_0(N)$, we write 
\begin{align}
  E_0(L, N) = F(L,N) + (-1)^N G(L,N),
\end{align}
where the functions $F$ and $G$ are smooth functions of their arguments, with $|G| \ll |F|$, as revealed by numerics in Fig. 1 (main text). The smooth  part can be decomposed into Fourier harmonics, each with a characteristic wave number. Taking a discrete derivative of $F$ amounts to multiplication of each mode by its characteristic frequency, effectively suppressing its contribution. In contrast, the effect of each successive discrete derivative on the second term is to amplify it by a factor of 2, yielding  
\begin{eqnarray}
     E^{(k)}_0(N) = 2^k (-1)^\frac{k}{2} G(L,N) (-1)^N + O(\kappa^k F)
\end{eqnarray}
for even derivatives.
The second discrete derivative, 
\begin{align}
    E^{(2)}_0(N) = E_0(N+1)+E_0(N-1)-2E_0(N),
\end{align}
evaluated at $L=4N$ (Fig. 1 in main text) appears smooth. In contrast, the fourth derivative, 
\begin{align}
\begin{split}
E^{(4)}_0(N) = E_0(N+2) - 4E_0(N+1) + 6E_0(N)\\ - 4E_0(N-1) +E_0(N-2),
\end{split}
\end{align}
(again evaluated at $L=4N$) reveals clear oscillations with $N$ (see Fig. \ref{fig:E^(4)_0}). The oscillations can be attributed to tiny oscillations in $E_0(N)$ originating from occupying and de-occupying the MZM mode as the parity of $N$ changes.
The magnitude of the `overlap energy,' $G(L, N) \sim |E^{(4)}_0(N)/2^4|$,  is indeed quite small even for our limited sizes accessible with DMRG. 

Of course, an even better approximation for the overlap energy could be obtained by taking more derivatives (e.g., $E^{(n)}_0(N)/2^n$), but $E^{(4)}_0(N)$ already exhibits clear sign oscillations, with $|E^{(4)}_0(N)|$ being nearly smooth. 

We can estimate the value of $G$ by isolating the energy difference between the odd- and even-parity states, using our knowledge of the correlation matrices.  Since the kinetic energy operator is spatially local, it is unimportant for evaluating this energy difference, which manifests in long-range correlators. We thus focus on the interaction energy
\begin{eqnarray}
    H_{int} = -\frac{4g}{L} \sum_{i>j, m>\ell}^L \eta(i-j)\eta(\ell-m)c^\dag_j c_i^\dag c_m c_\ell.
\end{eqnarray}
We seek the topological contribution to the energy difference. For estimation purposes, we assume that the state is approximately Gaussian (mean-field like), and thus can be Wick-decoupled in the normal and anomalous channels. 
The topological splitting then  comes from terms proportional to $M\Delta M$ and $P\Delta P$. The latter contributes to the odd-even splitting as

\begin{eqnarray}
    G(L, N) \approx  -\frac{4g}{L} \sum_{i>j}^L \eta(i-j)P^*_{ij}\, \sum_{ m>\ell}^L \eta(\ell-m)\Delta P_{m\ell}.
\end{eqnarray}
For estimation purposes, we use the fact that $\eta(r)$ scales as $(-1)^r/r$.
Then the first sum scales with the system size as $LP_{loc}$, while the second sum scales as $P_{nl}/L$. The  contribution from the regular correlation matrix, $M\Delta M$, is parametrically smaller (by $1/L$) due to the special form of the interaction, and hence can be neglected. 
Finally,
\begin{eqnarray}
    G(L,N)\sim g P_{loc}P_{nl}/L.
\end{eqnarray}
The $1/L$ scaling of the odd-even splitting is indeed consistent with the slow decay observed in Fig. \ref{fig:E^(4)_0} (remember, $L = 4N$). 
Quantitatively, using the values from Fig. \ref{fig:anomolous}, $P_{loc}\sim P_{nl}\sim 0.1$ for $L = 32, \ N = 8$,
$G(32, 8)\sim 10^{-2}/32\sim 3\times 10^{-4}$. This has the correct scale, comparable with the obtained $E^{(4)}_0/2^4$. 
We thus conclude that the slow decay of the MZM splitting in the RGK chain can be attributed to the long-range nature of the interactions.

\begin{figure}
\includegraphics[scale=0.40]{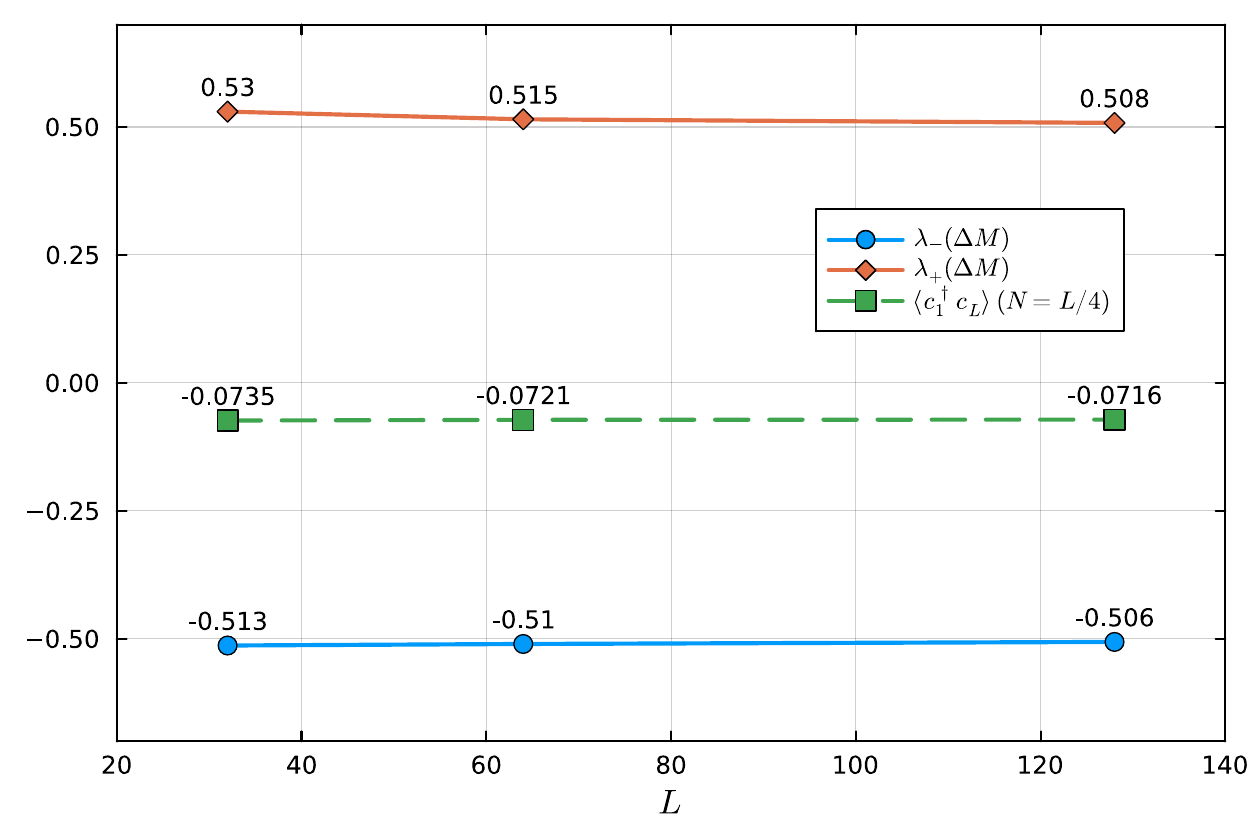}
\caption{Scaling behavior of the leading eigenvalues of $\Delta M$ and $\langle c_1^\dag c_L \rangle$ for $g=1$ (RGK model).} 
\label{fig:scaling}
\end{figure} 

\section{Convergence of DMRG}
Since our conclusions rely on the numerical accuracy of DMRG, we include this section to demonstrate the convergence of DMRG for some important quantities in the RGK model. We examine convergence not only with respect to system size, but also with respect to bond dimension and sweep count.

As discussed in the main text, the long range correlator $\langle c_1^\dag c_L\rangle$ saturates to a finite value as $L$ increases. We verify this remarkable behavior by computing $\langle c_1^\dag c_L\rangle$ up to $L=128$ (see Fig.~\ref{fig:scaling}). The correlator approaches a value of approximately $-0.07$ (for $N=L/4$), indicating that this effect is not a finite size artifact. For a given  system size $L$, we consider DMRG to be `well-converged' when: (i) the energy difference between the final two sweeps is less than $10^{-6}$; and (ii) increasing the maximum bond dimension for all sweeps changes the final energy by less than $10^{-6}$. We note that the number of sweeps required to satisfy these criteria increases with $L$.

We now turn to the behavior of $\Delta M$. As $L$ increases, the separation between the dominant eigenvalues and the sub-leading ones becomes more pronounced, consistent
with the expectation that the distinction between trivial and topological behavior becomes exact in
the thermodynamic limit. Conversely, for smaller systems, finite-size effects lead to less pronounced eigenvalue separation. This trend is illustrated in Fig.~\ref{fig:scaling}, which shows the leading eigenvalues of $\Delta M$ as a function of $L$ for $g = 1$.

\end{document}